\documentclass[10pt,a4paper,english]{article}
\usepackage{amsmath,amscd,amsfonts,eucal,latexsym,amssymb,mathrsfs}
\usepackage{amsxtra, amsthm, calc, bbm}
\newlength{\papersidemargin}
\newlength{\papertopmargin}
\newlength{\maxcarwidth}
\theoremstyle{definition}
\newtheorem{definition}{Definition}[section]
\theoremstyle{plain}
\newtheorem{theorem}[definition]{Theorem}
\newtheorem{proposition}[definition]{Proposition}
\newtheorem{lemma}[definition]{Lemma}
\newtheorem{corollary}[definition]{Corollary}

\theoremstyle{remark}
\newtheorem*{Remark}{Remark}
\newtheorem*{Remarks}{Remarks}
\numberwithin{equation}{section}
\newcounter{remcount}

\newenvironment{remlist}{\begin{list}{(\roman{remcount})}{\usecounter{remcount}\setlength{\leftmargin}{0 cm}\setlength{\rightmargin}{0 cm}\setlength{\topsep}{0 cm}\setlength{\itemsep}{0 cm}\setlength{\parsep}{\parskip}\setlength{\labelwidth}{1 cm}\setlength{\labelsep}{0.5 em}\setlength{\itemindent}{1 cm + 0.5 em}}}{\end{list}}
\newcounter{propcount}

\newlength{\maxlabelwidth}
\newenvironment{proplist}[2][1]{\begin{list}{(\roman{propcount})}{\usecounter{propcount}\setcounter{propcount}{#2}\settowidth{\maxlabelwidth}{\textit{(\roman{propcount})}}\setcounter{propcount}{#1-1}\setlength{\leftmargin}{\maxlabelwidth+ 
0.5 em}\setlength{\rightmargin}{0 cm}\setlength{\topsep}{0 
cm}\setlength{\itemsep}{0 
cm}\setlength{\parsep}{\parskip}\setlength{\labelwidth}{\maxlabelwidth}\setlength{\labelsep}{0.5 
em}\setlength{\itemindent}{0 cm}}}{\end{list}}
\def\bC{{\mathbb C}}

\def\bN{{\mathbb N}}

\def\bR{{\mathbb R}}

\def\bZ{{\mathbb Z}}
\def\a{\alpha}
\def\b{\beta}
        
\def\d{\delta}        \def\D{\Delta}
\def\eps{\varepsilon} 
     
\def\z{\zeta}
\def\e{\eta}
\def\th{\theta}    \def\Th{\Theta}
\def\k{\kappa}
\def\l{\lambda}       \def\L{\Lambda}
\def\m{\mu}
\def\n{\nu}
\def\x{\xi}
\def\p{\pi}
\def\r{\rho}
\def\s{\sigma}

\def\f{\varphi}

\def\o{\omega}        \def\O{\Omega}
\def\fA{{\mathfrak A}}

\def\fF{{\mathfrak F}}
\def\fG{{\mathfrak G}}

\def\fW{{\mathfrak W}}

\newcommand{\sA}{\mathscr{A}}
\newcommand{\sB}{\mathscr{B}}

\newcommand{\sD}{\mathscr{D}}

\newcommand{\sF}{\mathscr{F}}
\newcommand{\sG}{\mathscr{G}}
\newcommand{\sH}{\mathscr{H}}

\newcommand{\sP}{\mathscr{P}}

\newcommand{\sS}{\mathscr{S}}

\newcommand{\supp}{{\textup{supp}\,}}
\newcommand{\ad}{{\textup{Ad}\,}}
\newcommand{\rest}{\upharpoonright}

\newcommand{\Id}{\mathbbm{1}} 
\newcommand{\ueq}{\cong}
\newcommand{\iso}{\simeq}

\renewcommand{\Im}{\textup{Im}\,}
\newcommand{\vntensor}{\bar{\otimes}}
\newcommand{\ran}{\textup{ran}\,}
\newcommand{\abs}[1]{\lvert#1\rvert}

\newcommand{\norm}[1]{\| #1 \|}
\newcommand{\bignorm}[1]{\big\| #1 \big\|}
\newcommand{\Bignorm}[1]{\Big\| #1 \Big\|}
\newcommand{\scalar}[2]{\langle #1  ,  #2\rangle}
\newcommand{\bigscalar}[2]{\big\langle #1  ,  #2\big\rangle}

\DeclareMathOperator*{\sslim}{s*-lim}

\newcommand{\hO}{\hat{O}}
\newcommand{\Vp}{V_+}
\newcommand{\Vpc}{\overline{V}_+}
\newcommand{\Pg}{\sP}

\newcommand{\Pport}{\Pg_+^\uparrow}
\newcommand{\rPport}{\tilde{\Pg}_+^\uparrow}

\newcommand{\Lx}{{(\Lambda , x)}}                
\newcommand{\bp}{\boldsymbol{p}}
\newcommand{\bx}{\boldsymbol{x}}
\newcommand{\hV}{\hat{V}}
\newcommand{\sFi}{\sF^{(i)}}
\newcommand{\sFo}{\sF^{(1)}}
\newcommand{\sFt}{\sF^{(2)}}

\newcommand{\sAo}{\sA^{(1)}}

\newcommand{\Ui}{U^{(i)}}
\newcommand{\Uo}{U^{(1)}}
\newcommand{\Ut}{U^{(2)}}
\newcommand{\Vi}{V^{(i)}}
\newcommand{\Vo}{V^{(1)}}
\newcommand{\Vt}{V^{(2)}}
\newcommand{\Oi}{\O^{(i)}}
\newcommand{\Oo}{\O^{(1)}}
\newcommand{\Ot}{\O^{(2)}}

\newcommand{\sHo}{\sH^{(1)}}
\newcommand{\sHt}{\sH^{(2)}}
\newcommand{\ai}{\a^{(i)}}
\newcommand{\bi}{\b^{(i)}}
\newcommand{\oi}{\o^{(i)}}
\newcommand{\Ei}{E^{(i)}}
\newcommand{\ao}{\a^{(1)}}
\newcommand{\bo}{\b^{(1)}}
\newcommand{\oo}{\o^{(1)}}

\newcommand{\bt}{\b^{(2)}}
\newcommand{\ot}{\o^{(2)}}
\newcommand{\Et}{E^{(2)}}

\newcommand{\TbO}{\Th_{\b,O}}
\newcommand{\TlblO}{\Th_{\l \b,\l O}}
\newcommand{\TzbO}{\Th^{(0)}_{\b,O}}
\newcommand{\TzbhO}{\Th^{(0)}_{\b,\hO}}

\newcommand{\TfpO}{\Th_{f,\psi,O}}
\newcommand{\TifpO}{\Th^{(i)}_{f,\psi,O}}
\newcommand{\TofpO}{\Th^{(1)}_{f,\psi_1,O}}
\newcommand{\TtfpO}{\Th^{(2)}_{f,\psi_2,O}}
\newcommand{\TzfpO}{\Th^{(0)}_{f,\psi,O}}
\newcommand{\TzfphO}{\Th^{(0)}_{f,\psi,\hO}}
\newcommand{\TflplO}{\Th_{f^\l,\psi,\l O}}
\newcommand{\ToflplO}{\Th^{(1)}_{f^\l,\psi_1,\l O}}
\newcommand{\TtflplO}{\Th^{(2)}_{f^\l,\psi_2,\l O}}
\newcommand{\TflplhO}{\Th_{f^\l,\psi,\l \hO}}
\newcommand{\TflkplkO}{\Th_{f^{\l_\k},\psi,\l_\k O}}
\newcommand{\TiflkplkO}{\Th^{(i)}_{f^{\l_\k},\psi_i,\l_\k O}}
\newcommand{\TflkplkhO}{\Th_{f^{\l_\k},\psi,\l_\k \hO}}
\newcommand{\PlN}{P^{(\l)}_N}
\newcommand{\PolN}{P^{(1,\l)}_N}
\newcommand{\PtlN}{P^{(2,\l)}_N}
\newcommand{\PlkN}{P^{(\l_\k)}_N}
\newcommand{\PilkN}{P^{(i,\l_\k)}_N}

\newcommand{\TlN}{T^{(\l)}_N}
\newcommand{\TlkN}{T^{(\l_\k)}_N}
\newcommand{\RlN}{R^{(\l)}_N}
\newcommand{\RlkN}{R^{(\l_\k)}_N}
\newcommand{\Nk}{N_\k}
\newcommand{\Nz}{N_0}
\newcommand{\uF}{\underline{F}}
\newcommand{\uFi}{\uF^{(i)}}
\newcommand{\uFo}{\uF^{(1)}}
\newcommand{\uFt}{\uF^{(2)}}
\newcommand{\uG}{\underline{G}}
\newcommand{\uH}{\underline{H}}
\newcommand{\uA}{\underline{A}}

\newcommand{\uId}{\underline{\Id}}
\newcommand{\uFl}{\underline{F}_{\lambda}}

\newcommand{\ufF}{\underline{\mathfrak F}}
\newcommand{\ufFi}{\ufF^{(i)}}
\newcommand{\ufFo}{\ufF^{(1)}}
\newcommand{\ufFt}{\ufF^{(2)}}
\newcommand{\utfF}{\underline{\tilde{\mathfrak F}}}
\newcommand{\ufA}{\underline{\mathfrak A}}

\newcommand{\ua}{\underline{\alpha}}

\newcommand{\SLF}{\textup{SL}^{\sF}}

\newcommand{\uf}{\underline{\varphi}}
\newcommand{\uo}{\underline{\omega}}
\newcommand{\uoo}{\underline{\omega}_0}

\newcommand{\uolk}{\underline{\omega}_{\lambda_\k}}
\newcommand{\ub}{\smash{\underline{\beta}}}
\newcommand{\ubt}{\smash{\underline{\b}}^{(2)}}
\newcommand{\sFz}{\sF_0}
\newcommand{\sFiz}{\sF^{(i)}_0}
\newcommand{\sFoz}{\sF^{(1)}_0}
\newcommand{\sFtz}{\sF^{(2)}_0}

\newcommand{\pz}{\pi_0}
\newcommand{\poz}{\pi^{(1)}_0}
\newcommand{\ptz}{\pi^{(2)}_0}

\newcommand{\sHz}{{\sH}_0}
\newcommand{\sHoz}{{\sH}^{(1)}_0}
\newcommand{\sHtz}{{\sH}^{(2)}_0}

\newcommand{\Oz}{\Omega_0}
\newcommand{\Oiz}{\Omega^{(i)}_0}
\newcommand{\Ooz}{\Omega^{(1)}_0}
\newcommand{\Otz}{\Omega^{(2)}_0}

\newcommand{\bps}{\boldsymbol{\psi}}
\newcommand{\sAL}{\sA_L}
\newcommand{\Wt}{\tilde{W}}
\newcommand{\tfW}{\tilde{\fW}}

\newcommand{\sHuo}{\sH^{(0)}}
\newcommand{\sAm}{\sA^{(m)}}
\newcommand{\sAz}{\sA^{(0)}}
\newcommand{\fAm}{\fA^{(m)}}

\newcommand{\am}{\a^{(m)}}
\newcommand{\amLx}{\a^{(m)}_{(\L,x)}}

\newcommand{\ts}{\tilde{\s}}
\newcommand{\td}{\tilde{\d}}
\newcommand{\om}{\o^{(m)}}
\newcommand{\oz}{\o^{(0)}}
\newcommand{\pim}{\p^{(m)}}
\newcommand{\piz}{\p^{(0)}}
\newcommand{\uAm}{\ufA^{(m)}}
\newcommand{\tth}{\tilde{\th}}
\newcommand{\tV}{\tilde{V}}

\newcommand{\bm}{\boldsymbol{\m}}
\newcommand{\abm}{\a^{(\bm)}}
\newcommand{\Ubm}{U^{(\bm)}}
\newcommand{\Ubz}{U^{(\bze)}}
\newcommand{\bze}{\boldsymbol{0}}

\newcommand{\obm}{\o^{(\bm)}}
\newcommand{\obz}{\o^{(\bze)}}
\newcommand{\Obm}{\O^{(\bm)}}
\newcommand{\Obz}{\O^{(\bze)}}
\newcommand{\sFbm}{\sF^{(\bm)}}
\newcommand{\sFbz}{\sF^{(\bze)}}
\newcommand{\sAbm}{\sA^{(\bm)}}
\newcommand{\uobmo}{\uo^{(\bm)}_0}
\newcommand{\fAmz}{\fAm_0}
\newcommand{\sAmz}{\sAm_0}
\newcommand{\uFbm}{\ufF^{(\bm)}}
\newcommand{\uabm}{\ua^{(\bm)}}
\newcommand{\Ubmz}{U^{(\bm)}_0}
\newcommand{\Vtz}{\Vt_0}

\newcommand{\sFbmz}{\sFbm_0}

\newcommand{\uGbm}{\underline{\fG}^{(\bm)}}
\newcommand{\tuGbm}{\smash{\underline{\tilde{\fG}}}^{(\bm)}}
\newcommand{\sGbmz}{\sG^{(\bm)}_0}
\begin{document}
\noindent
\begin{center}
{\Large\textbf{Scaling Algebras and Superselection Sectors:\\[0.3\baselineskip]
                Study of a Class of Models}}
\vspace{1.5\baselineskip}\\
{\large Claudio D'Antoni${}^{a,}$\footnote{supported by
    MIUR, INdAM-GNAMPA, and the Network ``Quantum Spaces - Noncommutative Geometry'' HPRN-CT-2002-00280.},
   Gerardo Morsella${}^{b,1}$}
\vspace{\baselineskip}\\
                 {\small\textit{${}^a$\,Dipartimento di Matematica, Universit\`a di Roma ``Tor Vergata'',\\
                 Via della Ricerca Scientifica, I-00133 Roma, Italy} \\
                 \texttt{dantoni$@$mat.uniroma2.it}
\vspace{0.5\baselineskip}\\
                 \textit{${}^b$\,Istituto Nazionale d'Alta Matematica ``Francesco Severi'' and \\
                 Dipartimento di Matematica, Universit\`a di Roma ``La Sapienza'', \\
                 P.le Aldo Moro 2, I-00185 Roma, Italy} \\ 
                 \texttt{morsella$@$mat.uniroma1.it}}
\end{center}
\setcounter{footnote}{0}
\begin{abstract}
We analyse a class of quantum field theory models illustrating some of the possibilities that have emerged in the general study of the short distance properties of superselection sectors, performed in a previous paper (together with R. Verch). In particular, we show that for each pair $(G,N)$, with $G$ a compact Lie group and $N$ a closed normal subgroup, there is a net of observable algebras which has (a subset of) DHR sectors in 1-1 correspondence with classes of irreducible representations of $G$, and such that only the sectors corresponding to representations of $G/N$ are preserved in the scaling limit. In the way of achieving this result, we derive sufficient conditions under which the scaling limit of a tensor product theory coincides with the product of the scaling limit theories.
\end{abstract}
\section{Introduction}\label{sec:intro}
The scaling algebra concept has been introduced in~\cite{Buchholz:1995a}, in an attempt to make available, in the framework of the algebraic approach to quantum field theory~\cite{Haag:1996a}, the methods of the renormalization group, which have proved very useful in analysing the short distance behaviour of quantum field theory in the conventional approach. The elements of the scaling algebra are functions of a scaling parameter $\l > 0$ taking values in the algebra of local observables of the theory under consideration, any of such function representing the orbit $\l \to R_\l(A)$ of an arbitrary observable $A$ under a family $(R_\l)_{\l > 0}$ of renormalization group transformations, whose choice is only restricted by the requirement that such orbits have a ``phase space occupation'' which is independent of the scale $\l$, i.e.\ that the operators $R_\l(A)$ are localized in regions of radius proportional to $\l$ and have energy-momentum transfer proportional to $\l^{-1}$. The information about the short distance (or, equivalently, high energy) properties of the given theory (to which we will refer, from now on, as the \emph{underlying} theory), is then obtained by studying the vacuum expectation values of such functions in the $\l \to 0$ limit, and is encoded in a new net of local observables, called the \emph{scaling limit} of the underlying net.

One of the major achievements of these methods has been the formulation of an intrinsic notion of charge confinement~\cite{Buchholz:1996xk}, not suffering from the ambiguities of the conventional one, which relies on the assignment of a physical interpretation to the unobservable fields in terms of which the theory is described (whose choice is of course highly non-unique). According to this new confinement notion, the underlying theory describes confined charges if the corresponding scaling limit theory has superselection sectors\footnote{We refer the reader to~\cite{Haag:1996a,Roberts:1989ps} for a comprehensive account of the theory of superselections sectors.} which are not, at the same time, sectors of the underlying theory itself. An example of such situation is provided by the Schwinger model (massless QED in two spacetime dimensions), which has trivial superselection structure at finite scales, but whose scaling limit theory exhibits nontrivial sectors~\cite{Buchholz:1996xk,Buchholz:1998vu}.

In order for this concept to be applied to a general theory, one needs a canonical way of comparing the superselection structures of the underlying theory and of the scaling limit one. With this aim in mind, a general study of the short distance properties of charged fields and of superselection sectors -- of both DHR and BF types -- has been performed in~\cite{D'Antoni:2003ay} (see also~\cite{D'Antoni:2004a}), where the scaling algebra and scaling limit concepts are extended to the nets of charge carrying fields localized in double cones or in spacelike cones (depending on the kind of sector with which these fields are associated), and are then used to formulate a notion of ``charge preservation'' in the scaling limit. In such a way, the confined sectors of the underlying theory are identified with those sectors of the scaling limit theory which do not arise as limits of preserved sectors of the underlying theory~\cite{D'Antoni:2004a}. For the convenience of the reader, we will give an account of some of the main results of this work in section~\ref{sec:scalingfields} below.

In the present paper, we study a class of quantum field theory models which exhibit both preserved and non-preserved DHR sectors, therefore providing an illustration of the general analysis of~\cite{D'Antoni:2003ay}. More precisely, for each pair $(G,N)$ consisting of a compact Lie group $G$ and of a normal closed subgroup $N \subset G$, we construct a local net $\sA$, satisfying the standard assumptions, which has (a subset of) DHR sectors labelled by the equivalence classes of unitary irreducible representations of $G$, and such that precisely the sectors corresponding to representations which are trivial on $N$ (i.e.\ representations which factorize through $G/N$) are preserved according to~\cite{D'Antoni:2003ay}, cfr.\ theorem~\ref{thm:main}. Similarly to~\cite{Doplicher:2002cb}, the net $\sA$ is obtained as the fixed point net $\sA = \sF^G$ of a suitable field net $\sF$ which carries an action of $G$, and in turn $\sF$ is defined as a tensor product $\sF = \sF_1 \otimes \sF_2$, where $\sF_1$ is a net with trivial scaling limit constructed using results in~\cite{Lutz:1997a} and generated by fields which carry the charges corresponding to representations of $G$ which are nontrivial on $N$, while $\sF_2$ is a free field net which has $G/N$ as gauge group. The above mentioned result amounts then to showing that: (i) thanks to the fact that $\sF_1$ has trivial scaling limit, the scaling limit of $\sF$ coincides with the scaling limit $\sF_{2,0}$ of $\sF_2$, and that (ii) the scaling limit net $\sF_{2,0}$ again has $G/N$ as its gauge group, and the corresponding sectors all comply with the preservation condition formulated in~\cite{D'Antoni:2003ay}. In order to establish point (i), we will derive, in section~\ref{sec:tensor}, sufficient conditions under which the operations of scaling limit and of forming the tensor product of two theories can be interchanged. Not surprisingly, the main assumption which we employ is that of ``asymptotic nuclearity'', definition~\ref{def:asympnuclear}, which was formulated in~\cite{Buchholz:1996mx}, and which plays here a role in allowing to approximate functions in the scaling algebra of the tensor product theory by finite sums of ``simple tensors'' of the form $\l \to \uF_{1\l}\otimes \uF_{2\l}$, with the $\uF_i$ in the scaling algebras of the factor theories. The proof of point (ii) above is then obtained in section~\ref{sec:model} by combining this result about the scaling limit of product theories, with the computation of the scaling limit of the free scalar field in~\cite{Buchholz:1998vu}. Together with a result in~\cite{D'Antoni:2003ay}, this also implies that the equivalence of local and global intertwiners holds for any theory generated by a finite number of multiplets of free scalar fields of arbitrary masses transforming under irreducible representations of a compact gauge group, corollary~\ref{cor:equivalence}.

\section{Scaling algebras for charged fields and preservation of DHR sectors}\label{sec:scalingfields}
For the paper to be reasonably self-contained, and in order to establish our notations, we give in the present section an exposition of the main results of~\cite{D'Antoni:2003ay} concerning the short distance analysis of DHR superselection sectors. We refer the interested reader to the original paper for more details and discussions of definitions and results.

 By a \emph{quantum field theory with gauge action} (QFTGA in the following) we mean a quintuple $(\sF, U, V, \O, k)$, such that:
\begin{remlist}
\item $O \to \sF(O)$ is a net of von Neumann algebras on open double cones in Minkowski $d$-dimensional spacetime ($d=3,4$) acting irreducibily on a Hilbert space $\sH$ with scalar product $\scalar{\cdot}{\cdot}$;
\item $U$ is a unitary strongly continuous representation on $\sH$ of the translations group $\bR^d$, satisfying the spectrum condition, i.e.\ the spectrum of $U$ is contained in the closed forward light cone, and with respect to which the net $\sF$ is covariant
\begin{equation*}
U(x) \sF(O) U(x)^* = \sF(O+x),\qquad x \in \bR^d;
\end{equation*}
we set $\a_x := \ad U(x)$;
\item $V$ is a unitary strongly continuous representation on $\sH$ of a compact gauge group $G$, which acts locally on $\sF$
\begin{equation*}
V(g) \sF(O) V(g)^* = \sF(O),\qquad g \in G,
\end{equation*}
and which commutes with $U$; we set $\b_g := \ad V(g)$, and the subnet of $G$-fixed points 
\begin{equation*}
\sA(O) := \sF(O)^G := \{F\in\sF(O)\,:\, \b_g(F)=F\,\forall g \in G\}
\end{equation*}
is the net of observables determined by $(\sF,U,V,\O,k)$;
\item $\O \in \sH$ is the vacuum vector, i.e. is the unique translation invariant unit vector in $\sH$ and it is cyclic for the quasi-local algebra $\fF := \overline{\bigcup_O \sF(O)}$ (closure in the uniform topology on $B(\sH)$); $\O$ is also gauge invariant, and we denote by $\o := \scalar{\O}{(\cdot)\O}$ the vacuum state;
\item  $k \in Z(G)$, $k^2 = e$, is the element defining the $\bZ_2$ grading according to which elements in the quasi-local algebra $\fF$ satisfy normal commutation relations, i.e. with
\begin{equation*}
F_\pm := \frac{1}{2}(F \pm \b_k(F)), \qquad F \in \fF,
\end{equation*}
and with $F_i \in \sF(O_i)$, $i=1,2$, $O_1$ and $O_2$ spacelike separated, one has
\begin{equation*}
F_{1,+} F_{2,+} = F_{2,+} F_{1,+},\quad F_{1,+} F_{2,-} = F_{2,-} F_{1,+},\quad F_{1,-} F_{2,-} = -F_{2,-} F_{1,-}.
\end{equation*}
\end{remlist}
When there is no risk of confusion, we will indicate the QFTGA $(\sF, U, V, \O, k)$ simply by $\sF$.

For simplicity we assumed here that we are dealing only with translations covariant nets, but most of the results in the present and following sections also hold for Poincar\'e covariant nets, i.e.\ QFTGAs $(\sF, U, V, \O, k)$ for which $U$ is actually a unitary representation of the universal covering $\rPport$ of the proper orthocronous Poincar\'e group, such that $U(\L,x) \sF(O) U(\L,x)^* = \sF(\L O+x)$. The notation $\a_{(\L,x)} := \ad U(\L,x)$ will be used in this case also.

The scaling algebra associated to $\sF$ is defined in the following way. On the C$^*$-algebra $B(\bR_+,\fF)$ of all norm bounded functions $\l \in \bR_+ \to \uF_\l \in \fF$, with the natural C$^*$-norm $\norm{\uF} = \sup_{\l > 0} \norm{\uF_\l}$, we define automorphic actions $\ua$ of $\bR^d$ and $\ub$ of $G$ by
\begin{equation*}
\ua_x(\uF)_\l := \a_{\l x}(\uFl),\quad \ub_g(\uF)_\l := \b_g(\uF_\l), \qquad x \in \bR^4, g \in G, \l > 0.
\end{equation*}
The \emph{local scaling algebra} of the double cone $O$ is then the C$^*$-algebra $\ufF(O)$ of all the functions $\uF \in B(\bR_+,\fF)$ such that $\uF_\l \in \sF(\l O)$ for each $\l > 0$, and
\begin{equation}\label{eq:cont}
\lim_{x \to 0} \norm{\ua_x(\uF) - \uF} = 0, \qquad \lim_{g \to e} \norm{\ub_g(\uF) - \uF} = 0.
\end{equation}
We will denote by $\ufF$ both the net $O \to \ufF(O)$ and the associated quasi-local C$^*$-algebra. 

Let $\f$ be a locally normal state of $\sF$, and define a net of states $(\uf_\l)_{\l > 0}$ on $\ufF$ by $\uf_\l(\uF) := \f(\uF_\l)$. We denote by $\SLF(\f)$ the set of weak$^*$ limit points of $(\uf_\l)_{\l > 0}$ for $\l \to 0$. From an argument due to Roberts~\cite{Roberts:1974aa}, it follows that for any pair $\f_1$, $\f_2$ of locally normal states on $\sF$, there holds
\begin{equation}\label{eq:roberts}
\lim_{\l \to 0} \bignorm{(\f_1 - \f_2)\rest \sF(\l O)}= 0,
\end{equation}
and then $\SLF(\f)$ is actually independent of $\f$, and is called the set of \emph{scaling limit states} of $\ufF$. It easily follows that any $\uoo \in \SLF$ is $\ua$- and $\ub$-invariant and then, if $(\pz,\sHz,\Oz)$ is the corresponding GNS representation, by defining the net of von Neumann algebras
\begin{equation*}
\sFz(O) := \pz(\ufF(O))'',
\end{equation*}
and the representations $U_0$ of $\bR^d$ and $V_0$ of $G_0 := G/N_0$ (where $N_0 := \{ g \in G\, :\, \pz(\ub_g(\uF)-\uF)\Oz = 0,\, \forall \uF \in \ufF\}$) by
\begin{equation*}
U_0(x)\pz(\uF)\Oz := \pz(\ua_x(\uF))\Oz, \qquad V_0(gN_0)\pz(\uF)\Oz:= \pz(\ub_g(\uF))\Oz,
\end{equation*}
one gets that $(\sFz,U_0,V_0,\Oz,kN_0)$ is a QFTGA such that $\sA_0(O):= \pz(\ufA(O))'' = \sFz(O)^{G_0}$, $\ufA$ being the scaling algebra for the observable net $\sA$ defined in~\cite{Buchholz:1995a}.

\begin{Remark} We note, for future reference, that if the net $\sF$ is Poincar\'e covariant, then also the nets $\ufF$ and $\sFz$ can be made Poincar\'e covariant by extending $\ua$ and $U_0$ to $\rPport$ by
\begin{equation*}
\ua_{(\L,x)}(\uF)_\l := \a_{(\L,\l x)}(\uF_\l), \qquad U_0(\L,x)\pz(\uF)\Oz := \pz(\ua_{(\L,x)}(\uF))\Oz,
\end{equation*}
but, in general, the function $\L \to U_0(\L,x)$ will not be strongly continuous, since, at variance with what is done in~\cite{D'Antoni:2003ay}, we are not requiring here that condition~\eqref{eq:cont} is satisfied for the extended $\ua$.
\end{Remark}

If we now assume that $\sF$ is the covariant field net arising from a net of local observables $\sA$ through the Doplicher-Roberts reconstruction theorem~\cite{Doplicher:1990a}, we can define the notion of preservation of DHR sectors in the scaling limit. We first recall that to any (finite statistics, covariant) sector $\x$ of $\sA$ we can associate, for any double cone $O$, a \emph{multiplet of class} $\x$ of field operators, i.e.\ elements $\psi_j \in \sF(O)$, $j=1,\dots,d$, with $d$ the statistical dimension of $\x$, such that
\begin{equation*}
\psi_i^* \psi_j = \d_{ij}\Id, \qquad \sum_{j=1}^d \psi_j \psi_j^* = \Id, \qquad \b_g(\psi_i) = \sum_{j=1}^d \psi_j v_\x(g)_{ji},
\end{equation*}
where $v_\x$ is a unitary irreducible representation of $G$ in the class associated to the sector $\x$.

We will then say that a finite statistics, covariant sector $\x$ of $\sA$ is \emph{preserved in the scaling limit state} $\uoo$ if for each double cone $O_1$ and each $\l > 0$ it is possible to find a multiplet of class $\x$, $\psi_j(\l) \in \sF(\l O_1)$, $j=1,\dots,d$, such that for each $\eps > 0$, each double cone $O$ containing the closure of $O_1$ and each $j=1,\dots,d$, there exist scaling algebra elements $\uF, \uF' \in \ufF(O)$ for which
\begin{equation}\label{eq:preserv}
\limsup_{\k} \big(\norm{[\psi_j(\l_\k)-\uF_{\l_\k}]\O}+\norm{[\psi_j(\l_\k)-\uF'_{\l_\k}]^*\O} \big)< \eps,
\end{equation}
where $(\l_\k)_{\k \in K} \subset \bR_+$ is a net such that $\uoo = \lim_\k \uo_{\l_\k}$.

As discussed at length in~\cite{D'Antoni:2003ay}, the restriction that the above condition imposes on the sector $\x$ is essentially that the states $\psi_j(\l)\O$, which represent a charge $\x$ roughly localized in the region $\l O$, should have energy-momentum scaling not faster than $\l^{-1}$, and this corresponds to the physical picture that a preserved charge should be ``pointlike'', and therefore its phase space occupation should only be restricted by Heisenberg principle, as opposed to a charge with some ``internal structure'' which requires a surplus of energy in order to be localized in small regions.

In order to state the consequences of such notion of charge preservation, we introduce here a notation which will also be useful in the following. For a bounded function $\l \in \bR_+ \to F_\l \in \sF(\l O)$ 
and functions $h \in L^1(\bR^d)$, $\psi \in L^1(G)$, we set
\begin{equation}
(\ua_hF)_\l := \int_{\bR^d}dx \,h(x) \a_{\l x}(F_\l), \quad (\ub_\psi F)_\l := \int_G dg \,\psi(g) \b_g(F_\l),
\end{equation}
where $dg$ is the normalized Haar measure on $G$ and the integrals are understood in weak sense. It is easy to verify that $G_\l := (\ub_\psi F)_\l$ is such that 
\begin{equation}\label{eq:gaugecont}
\lim_{g \to e} \sup_{\l > 0}\norm{\b_g(G_\l)-G_\l} = 0,
\end{equation}
and for any function $\l \to G_\l$ satisfying this condition $\ua_h G \in \ufF$, and $\ua_h G \in \ufF(\hat{O})$ if $\hat{O} \supset O + \supp h$.

If the sector $\x$ is preserved in the state $\uoo$ and $\psi_j(\l)$ is a multiplet satisfying~\eqref{eq:preserv}, we obtain that for each $\d$-sequence $(h_n)_{n\in\bN}$ the limit
\begin{equation*}
\bps_j := \sslim_{n\to +\infty}\pz(\ua_{h_n}\psi_j),
\end{equation*}
exists in the strong$^*$ operator topology, is independent of the chosen $\d$-sequence, and defines a multiplet of class $\x$ in $\sFz(O)$ (in the sense that the representation $v_\x$ is trivial on $N_0$, and $\bps_j$ is a multiplet of the corresponding representation of $G_0$), and furthermore, by defining
\begin{equation*}
\boldsymbol{\rho}(A) := \sum_{j=1}^d \bps_j A \bps_j^*, \qquad A \in \fA_0,
\end{equation*}
with $\fA_0$ the quasi-local algebra of the net $\sA_0$, one gets a DHR endomorphism of $\fA_0$, whose sector is therefore identified with the scaling limit of the sector $\x$.

The last result that we cite from~\cite{D'Antoni:2003ay} is the following generalization of a theorem proven by Roberts~\cite{Roberts:1974aa} for dilatation invariant theories: if all the sectors of the underlying theory are preserved in some scaling limit state, and if the local field algebras are factors, $\sF(O)\cap\sF(O)' = \bC\Id$, then local intertwiners between DHR endomorphisms of $\fA$ are also global intertwiners, i.e.\ if $\rho,\s$ are covariant, finite statistics DHR endomorphisms of $\fA$ and $T \in \fA$ is such that $T\rho(A) = \s(A)T$ holds for each $A \in \sA(O)$, then it also holds for each $A \in \fA$.

\section{Scaling limit of tensor product theories}\label{sec:tensor}
As mentioned in the introduction, nuclearity assumptions will play a fundamental role in the discussion of the scaling limit of tensor product theories. For the notion of a $p$-nuclear map between Banach spaces, see definition~\ref{def:nuclear} in appendix~\ref{app:nuclear}. 

Let $(\sF,U,V,\O,k)$ be a QFTGA. For a non-negative function $\psi \in  C(G)$ with $\int_G \psi = 1$ we introduce the notation
\begin{equation*}
\hV(\psi) := \int_G dg\,\psi(g) V(g),
\end{equation*}
where the integral is defined in the strong sense. Of course $\norm{\hV(\psi)}= 1$. For a double cone $O$ and a function $f \in C_b(\bR^d)$, consider the map $\TfpO : \sF(O) \to \sH$ defined by
\begin{equation}\label{eq:gennuclearmap}
\TfpO(F) := f(P)\hV(\psi)F\O, \qquad F \in \sF(O),
\end{equation}
where $P$ is the $d$-momentum operator of our theory, i.e.\ the generator of the translations group. An important particular case of such maps is the map $\TbO$ obtained when $\psi$ approaches a $\d$-function at $e\in G$ and $f$ is such that $f(p) = e^{-\b p_0}$ for $p \in \overline{V}_+$, for some $\b > 0$, i.e.
\begin{equation}\label{eq:nuclearmap}
\TbO(F) := e^{-\beta H}F\O, \qquad F \in \sF(O),
\end{equation}
with $H = P_0$ the generator of time translations.

\begin{definition}\label{def:asympnuclear}
The QFTGA $(\sF,U,V,\O,k)$ is said to be \emph{asymptotically (uniformly) $p$-nuclear} if all the maps $\TbO$ are $p$-nuclear and
\begin{equation}\label{eq:asympnuclear}
\limsup_{\l \to 0} \norm{\TlblO}_p < +\infty.
\end{equation}
\end{definition}

From the estimates in~\cite[prop. 3.1]{Mohrdieck:2002a}, it follows that the theory of $n$ free scalar fields of masses $m_i \geq 0$, $i=1,\dots, n$, is asymptotically $p$-nuclear for any $p \in (0,1]$.

The notion of asymptotic nuclearity was first introduced
in~\cite{Buchholz:1996mx}, were the relations between the phase space
properties of the underlying theory and the structure of its scaling
limits were analysed. Essentially all the results to be found there
can be generalized to the present setting (the generalization consisting in the fact that we here allow for a nontrivial gauge group $G$ acting on the net, as well as for normal commutation relations). 

In particular, we will need the following results, whose proofs are
obtained by a straightforward modification of the ones of theorems 4.5
and 4.6 in~\cite{Buchholz:1996mx}, combined with the
remark~\cite[lemma 3.1]{Mohrdieck:2002a} that the nuclearity
properties of the map $\TzbO$, defined as the analogue of the map
$\TbO$ for a given scaling limit theory $\sFz$, are the same as the ones of the map $F \in \pz(\ufF(O)) \to e^{-\b H_0}F\Oz$ which is considered in~\cite{Buchholz:1996mx} (we recall that $\sFz(O) = \pz(\ufF(O))^-$).

\begin{proposition}\label{prop:nuclearlimit}
Assume that the theory $\sF$ is asymptotically $p$-nuclear for $0 < p < 1/3$. Then:
\begin{proplist}{1}
\item for each scaling limit theory $\sFz$ the corresponding maps $\TzbO$ are $q$-nuclear for any $q > 2p/(2-3p)$, and there exists a $c > 0$, depending only on $p$, $q$, such that
\begin{equation*}
\norm{\TzbO}_q \leq c \limsup_{\l \to 0} \norm{\TlblO}_p;
\end{equation*}
\item \label{part:classical} if there exists a constant $c$ such that
\begin{equation*}
\limsup_{\l \to 0} \norm{\TlblO}_p\leq c,
\end{equation*}
uniformly for all double cones $O$, then $\sF$ has a classical scaling limit.
\end{proplist}
\end{proposition} 

We now state and prove some technical results that we will use later
in the discussion of the scaling limit of a tensor product theory. 

We first introduce some notation. For $f \in \sS(\bR^d)$ we adopt the following conventions for its
Fourier transform and anti-transform:
\begin{equation*}
\hat{f}(p) := \int_{\bR^d} dx\, f(x)e^{ipx}, \quad \check{f}(x) := \int_{\bR^d} \frac{dp}{(2\pi)^d}f(p)e^{-ipx},
\end{equation*}
where of course $px = p_\mu x^\mu$ is the Minkowski scalar product of $p,x \in \bR^d$. Also, for a function $f$ on $\bR^d$, and $\l > 0$, we set $f^\l(p) = f(\l p)$, $p \in \bR^d$.

\begin{lemma}\label{lem:sSfO}
Let the theory $\sF$ be asymptotically $p$-nuclear for $0 < p < 1/3$, let $\uoo$ be a scaling limit state and $f \in \sS(\bR^d)$ be such that $\sup_{p \in \overline{V}_+} \abs{f(p)e^{\b p_0}} < \infty$ for some $\b > 0$. Then if 
$2p/(1-p) < q \leq 1$, for each double cone $\hat{O}$ and for each $\eps > 0$ there are elements $\uF_1, \dots, \uF_N \in \ufF(\hat{O})$ such that, if we let
\begin{equation}\label{eq:PN}
P_N := \sum_{n=1}^N \bigscalar{\TzfphO(\pz(\uF_n))}{\cdot}\TzfphO(\pz(\uF_n)),
\end{equation}
then
\begin{equation}\label{eq:idminuspn}
\norm{(\Id-P_N)\TzfphO}_q < \eps.
\end{equation}
\end{lemma}

\begin{proof}
From the conditions on $p, q$ in the statement, it follows that we can take a number $r$ such that $2p/(2-3p) < r < 2q/(4-q)$, which implies that $q > 4r/(r+2)$ and $r < 2/3$. This implies, according to the previous proposition, that $\TzbhO$ is $r$-nuclear, and then, since it follows from the conditions on the function $f$ that $f(P)e^{\b H}$ is a bounded operator on $\sHz$, $\TzfphO = f(P)e^{\b H}\hV_0(\psi)\TzbhO$ is $r$-nuclear too. Then, according to lemma~\ref{lem:nuclearbon} in appendix~\ref{app:nuclear}, there exist an orthonormal system $(\Phi_n)_{n \in \bN} \subset \overline{\ran \TzfphO}$ and a family $(\f_n)_{n\in\bN} \subset \sFz(\hat{O})^*$ such that
\begin{equation*}
\TzfphO(F) = \sum_{n=1}^{+\infty}\f_n(F)\Phi_n, \qquad \sum_{n=1}^{+\infty}\norm{\f_n}^q <+\infty.
\end{equation*}
It is therefore possible to find an integer $N$ such that if $Q_N$ is the orthogonal projection on the subspace spanned by $\Phi_1,\dots,\Phi_N$,
\begin{equation}\label{eq:idminusqn}
\norm{(\Id-Q_N)\TzfphO}_q \leq \bigg(\sum_{n=N+1}^{+\infty}\norm{\f_n}^q\bigg)^{1/q} < \frac{\eps}{2}.
\end{equation}
Furthermore, since, as it is easily checked, $\overline{\ran \TzfphO} = \overline{\TzfphO(\pz(\ufF(\hO)))}$, we can find elements 
$\uF_n \in \ufF(\hO)$, $n = 1, \dots, N$, such that
\begin{equation*}
\norm{\Phi_n - \TzfphO(\pz(\uF_n))} < \min\bigg\{1,\frac{\eps}{3\cdot 2^{n+1}\norm{\TzfphO}_q}\bigg\},
\end{equation*}
so that $\norm{\TzfphO(\pz(\uF_n))} \leq 2$. Then, if $P_N$ is given by
equation~\eqref{eq:PN}, we get, for each $\Phi \in \sHz$,
\begin{equation*}\begin{split}
\norm{(Q_N-P_N)\Phi} &\leq \sum_{n=1}^N\bignorm{\scalar{\Phi_n}{\Phi}\Phi_n - \bigscalar{\TzfphO(\pz(\uF_n))}{\Phi}\TzfphO(\pz(\uF_n))} \\
                     &\leq 3\sum_{n=1}^N\norm{\Phi_n - \TzfphO(\pz(\uF_n))}\norm{\Phi} < \frac{\eps}{2\norm{\TzfphO}_q}\norm{\Phi},
\end{split}\end{equation*}
i.e.\ $\norm{Q_N-P_N}\leq \eps/2\norm{\TzfphO}_q$, which, together with inequality~\eqref{eq:idminusqn}, gives the statement.
\end{proof} 

In order not to burden the formulas too much, in the following lemma and in the proof of lemma~\ref{lem:approxtildeF}, we will make the following slight abuse of notation: given an element $\uF_n \in \ufF$ we will denote its value at scale $\l$ as $\uF_{n\l}$ (instead of $(\uF_n)_\l$), which should not be confused with the value of an element $\uF$ at scale $n\l$.

\begin{lemma}\label{lem:approxtheta}
Assume that the theory $\sF$ is asymptotically $p$-nuclear for $p \in (0,1/6)$, and let $f \in \sS(\bR^d)$ be as in the previous lemma, and $\uoo = \lim_{\k \in K}\uolk$ be a scaling limit state of $\sF$. Then if $2p/(1-4p)<q\leq 1$, for each pair of double cones $O, \hO$ with $\overline{O} \subset \hO$ and for each $\eps > 0$ there exist $\uF_1, \dots, \uF_N \in \ufF(\hO)$ such that, if we set
\begin{equation}\label{eq:PlN}
\PlN := \sum_{n=1}^N\bigscalar{\TflplhO(\uF_{n\l})}{\cdot}\TflplhO(\uF_{n\l}),
\end{equation}
we have
\begin{equation}\label{eq:tminuspnt}
\limsup_{\k \in K} \bignorm{\TflkplkO - \PlkN \TflkplkO}_q \leq \eps,
\end{equation}
where for each $\k \in K$ the $q$-norm appearing in the last equation is the $q$-norm of nuclear maps in $\sB(\sF(\l_\k O),\sH)$.
\end{lemma}

\begin{proof}
We use a variation of the arguments in~\cite[thm. 4.5]{Buchholz:1996mx}. We observe preliminarly that given bounded functions $\l \to F_\l \in \sF(\l O_1)$, $\l \to G_\l \in \sF(\l O_2)$ we have
\begin{multline}\label{eq:limscalar}
\lim_\k \bigscalar{\Th_{f^{\l_\k},\psi,\l_\k O_1}(F_{\l_\k})}{\Th_{f^{\l_\k},\psi,\l_\k O_2}(G_{\l_\k})} 
\\= \lim_\k \bigscalar{(\ua_{\check{f}}\ub_\psi F)_{\l_\k}\O}{(\ua_{\check{f}}\ub_\psi G)_{\l_\k} \O} = \bigscalar{\pz(\ua_{\check{f}}\ub_\psi F)\Oz}{\pz(\ua_{\check{f}}\ub_\psi G)\Oz}.
\end{multline}

For simplicity, for any given family $\uF_n \in \ufF(\hO)$, $n=1,\dots,N$,
and for the corresponding $\PlN$ defined as in~\eqref{eq:PlN}, we set $\TlN := \TflplO - \PlN\TflplO$. Furthermore, we denote by $\Nk(\eps)$ the $\eps$-content of the map $\TlkN$, and by $\Nz(\eps)$ that of the map $(\Id-P_N)\TzfphO$, where $P_N$ is defined as in the previous lemma, equation~\eqref{eq:PN}.

We begin by showing that the following inequality holds for each $\eps>0$:
\begin{equation}\label{eq:inequalepscontent}
\limsup_\k \Nk(\eps) \leq \Nz(\eps/2).
\end{equation} 
If this is not true, there exists an $\eps > 0$ such that, if we set $M := \Nz(\eps/2)$, for each $\n \in K$ we can find a $\k(\n) \in K$, $\k(\n) \geq \n$, and elements  $G^{(n)}_\n \in \sF(\l_{\k(\n)}O)$, $\norm{G^{(n)}_\n} \leq 1$, $n=1,\dots, M+1$, such that
\begin{equation*}
\norm{T^{(\l_{\k(\n)})}_N(G^{(n)}_\n-G^{(m)}_\n)} > \eps
\end{equation*}
if $n\neq m$.
Define then, for each $\l > 0$, and $n=1,\dots,M+1$,
\begin{equation*}
G^{(n)}_\l := \begin{cases}G^{(n)}_\n \quad &\text{if $\l = \l_{\k(\n)}$ for some $\n \in K$}, \\
                           0            &\text{otherwise.} \end{cases}
\end{equation*}
It is straightforward to check that the set $\tilde{K} := \{ \k(\n)\, : \, \n \in K\} \subset K$ is, with the induced partial ordering, a subnet of $K$, and therefore it is easy to verify, using~\eqref{eq:limscalar}, that
\begin{multline}\label{eq:inequalinterm}
\bignorm{(\Id-P_N)\pz\big(\ua_{\check{f}}\ub_\psi G^{(n)}-\ua_{\check{f}}\ub_\psi G^{(m)}\big)\Oz} \\= \lim_{\k \in K}\norm{\TlkN(G^{(n)}_{\l_\k}-G^{(m)}_{\l_\k})} 
= \lim_{\k \in \tilde{K}}\norm{\TlkN(G^{(n)}_{\l_\k}-G^{(m)}_{\l_\k})} \geq \eps.
\end{multline}
Pick now non-negative functions $h\in C_c(\bR^d),\chi \in C(G)$ with $\int_{\bR^d}h=1=\int_G \chi$ and $O + \supp h \subset \hO$, and define $\uH^{(n)} := \ua_h \ub_\chi G^{(n)} \in \ufF(\hO)$. Taking into account that convolution on $\bR^d$ is commutative and $G$ is unimodular, we see that we can take $\supp h$ and $\supp \chi$ so small that 
\begin{equation*}\begin{split}
\bignorm{&\TzfphO\big(\pz(\uH^{(n)})\big)-\pz\big(\ua_{\check{f}}\ub_\psi G^{(n)}\big)\Oz} \leq \norm{\ua_h\ua_{\check{f}}\ub_{\psi*\chi}G^{(n)} - \ua_{\check{f}} \ub_\psi G^{(n)}}\\
&\quad\leq \norm{\check{f}}_1 \sup_{g\in\supp \chi}\norm{\ub_{\psi_{g^{-1}}}G^{(n)} - \ub_\psi G^{(n)}}+\sup_{x\in\supp h}\norm{\ua_{\check{f}_x}\ub_\psi G^{(n)}-  \ua_{\check{f}}\ub_\psi G^{(n)}}  \\
&\quad< \frac{\eps}{4\norm{\Id-P_N}},
\end{split}\end{equation*}
where we used the standard notation $\check{f}_x(y) := \check{f}(y-x)$, $\psi_{g^{-1}}(h) := \psi(h g^{-1})$. Therefore, together with equation~\eqref{eq:inequalinterm}, we get
\begin{equation*}
\bignorm{(\Id-P_N)\TzfphO\big(\pz(\uH^{(n)})\big)-(\Id-P_N)\TzfphO\big(\pz(\uH^{(m)})\big)}>\eps/2,
\end{equation*}
which means that $N_0(\eps/2) \geq M + 1 = N_0(\eps/2)+1$, and this contradiction proves \eqref{eq:inequalepscontent}.

Now, according to lemma~\ref{lem:contentnuclear}.\ref{part:contenttonuclear} in appendix~\ref{app:nuclear}, there holds
\begin{equation}\label{eq:inequalnorm}
\limsup_\k \norm{\TlkN}_q \leq \limsup_\k \,d_q \bigg(\sum_{m=1}^{+\infty}(m^\frac{1}{2}\eps_m \Nk(\eps_m)^\frac{1}{m})^q\bigg)^\frac{1}{q},
\end{equation}
provided we can find a sequence of positive numbers $(\eps_m)_{m\in\bN}$ such that the series on the right hand side of this equation are convergent. We pick then numbers $r, s$ with $2p/(1-p) < r < 2q/(3q+2)$ and $r/(1-r)< s < 2q/(q+2)$ (that this is possible, follows from the conditions imposed on $p$, $q$) and set $\eps_m := \norm{(\Id-P_N)\TzfphO}_r m^{-\frac{1}{s}}$. It then follows from lemma~\ref{lem:contentnuclear}.\ref{part:nucleartocontent} that
\begin{equation*}
\eps_m \Nk(\eps_m)^\frac{1}{m} \leq \norm{(\Id-P_N)\TzfphO}_r \exp\bigg(\frac{c\norm{\TlkN}_r^s}{\norm{(\Id-P_N)\TzfphO}_r^s}\bigg)m^{-\frac{1}{s}},
\end{equation*}
and since, if $M:= \sup_{p \in \overline{V}_+} \abs{f(p)e^{\b p_0}}$, it is easily checked that
\begin{equation*}
\norm{\TlkN}_r \leq M\bigg(1+M^2\sum_{n=1}^N\norm{\uF_n}^2\bigg)\norm{\Th_{\l_\k\b,\l_\k O}}_r,
\end{equation*}
we have, from the assumption of asymptotic $p$-nuclearity of the theory, and from the fact that $r > p$ for $p\in(0,1/6)$, that there exists a constant $C>0$ and some $\n \in K$, independent of $m$, such that for all $\k > \n$ there holds 
\begin{equation*}
\eps_m \Nk(\eps_m)^\frac{1}{m} \leq C m^{-\frac{1}{s}}.
\end{equation*}
It follows from the conditions imposed on $q$, $s$ that the series $\sum_{m=1}^{+\infty}m^{\frac{q}{2}-\frac{q}{s}}$ is convergent, and we can then interchange the sum and the limit superior on the right hand side of~\eqref{eq:inequalnorm} obtaining a larger upper bound on the left hand side, so that, using inequality~\eqref{eq:inequalepscontent} and lemma~\ref{lem:contentnuclear}.\ref{part:nucleartocontent} once more, we conclude that there exists a constant $K_{q,s} > 0$ such that
\begin{equation*}
\limsup_\k \norm{\TlkN}_q \leq K_{q,s}\norm{(\Id-P_N)\TzfphO}_r,
\end{equation*}
and the statement is finally obtained by appealing to the previous lemma.
\end{proof}

We now pass to consider the situation in which we have two different
QFTGAs $(\sFi,\Ui,\Vi,\Oi,k_i)$, $i=1,2$. For simplicity we will
assume, in all what follows, that the $\sFi$ are purely bosonic, i.e.\
$k_i = e_i$ (identity of the group $G_i$). It is
straightforward, if cumbersome, to generalize the following results to
the case of two genuinely $\bZ_2$-graded nets (see the remarks after
theorem~\ref{thm:tensor}), but, as in the rest of the paper we will
need only the present special case, we refrain from giving details. Of course by defining
\begin{alignat}{2}
\sF(O) &:= \sFo(O)\vntensor\sFt(O),&\quad&\label{eq:tensorF}\\
U(x) &:= \Uo(x)\otimes\Ut(x), & &x \in \bR^d, \\
V(g_1,g_2) &:= \Vo(g_1)\otimes\Vt(g_2), &&(g_1,g_2) \in G_1\times G_2, \\
\O &:= \Oo\otimes\Ot,
\end{alignat}
we get a new QFTGA $(\sF,U,V,\O,(e_1,e_2))$ on the Hilbert space $\sH := \sHo \otimes \sHt$, which will be called the \emph{tensor product theory of $\sFo$ and $\sFt$}, and denoted, for brevity, with $\sFo\vntensor\sFt$. Our purpose is to study the relationship between the scaling limit theory of $\sF$ and the tensor product of the scaling limit theories of $\sFi$, $i=1,2$.

We recall that two QFTGAs $(\sF,U,V,\O,k)$ and $(\tilde{\sF},\tilde{U},\tilde{V},\tilde{\O},k)$ with the same gauge group are \emph{net-isomorphic} if there is an isomorphism of the quasi-local algebras $\th : \fF \to \tilde{\fF}$ such that $\th(\sF(O)) = \tilde{\sF}(O)$, $\tilde{\a}_x\th = \th\a_x$, $\tilde{\b}_g\th = \th\b_g$ and $\tilde{\o}\th = \o$, with obvious meaning of the symbols. It is then plain that the sets of scaling limit states of two net-isomorphic theories are in bijective correspondence, and that the scaling limit theories arising from two corresponding scaling limit states are net-isomorphic. Therefore net-isomorphic theories can be identified when discussing properties of their scaling limit theories. In particular, in the following we will always identify without further comment the nets $O \to \sFo(O)$ and $O \to \sFo(O)\vntensor\bC \Id \subset \sF(O)$, and the nets $O \to \sFt(O)$ and $O \to \bC\Id\vntensor \sFt(O) \subset \sF(O)$ (with the obvious definitions of translations and gauge transformations).

We will then denote by $\ufFi$ the scaling algebra associated to
$\sFi$, $i=1,2$. For $\uFi \in \ufFi(O)$, $i=1,2$, we define, by a
slight abuse of notation, $(\uFo\otimes\uFt)_\l :=
\uFo_\l\otimes\uFt_\l \in \sF(\l O)$, and it is clear that
$\uFo\otimes\uFt \in \ufF(O)$. We will denote by $\utfF(O)$ the
C$^*$-subalgebra of $\ufF(O)$ generated by such elements, and by
$\utfF$ the corresponding quasi-local C$^*$-algebra. We also define $\uId_\l := \Id$ for all $\l > 0$.

\begin{proposition}\label{prop:correspondence}
The sets of scaling limit states of the three theories $\sF$, $\sFo$, $\sFt$ are in bijective correspondence, in such a way that $\uoo \in \SLF(\o)$ corresponds to the states $\uF \in \ufFo \to \uoo(\uF\otimes\uId)$ in $\textup{SL}^{\sF^{(1)}}(\oo)$ and $\uF \in \ufFt \to \uoo(\uId\otimes\uF)$ in $\textup{SL}^{\sF^{(2)}}(\ot)$.
\end{proposition}

\begin{proof}
It is well known that there exist conditional expectations $\Ei : \fF
\to \fF^{(i)}$ such that $\Ei(\sF(O)) = \sFi(O)$, defined by the fact
that, say, $\Et(F)$, $F \in \sF(O)$, is the unique element of $\sFt(O)$ such that
$\phi(\Et(F)) = \oo\otimes\phi(F)$ for each $\phi \in \sFt(O)_*$, so
that $\Et(F_1\otimes F_2) = \oo(F_1)F_2$ (see for
instance the proof of thm.\ 2.6.4 in~\cite{Sakai:1971a}). It is then straightforward
to check that $\ai_x\Ei = \Ei\a_x$, $\bi_{g_i} \Ei = \Ei
\b_{(g_1,g_2)}$ and $\oi\Ei = \o$. It can then be
shown~\cite{Conti:2004a} that given a conditional expectation between
two nets with the above properties, the respective sets of scaling
limit states are in bijective correspondence, such correspondence
being given by the restriction of scaling limit states.
\end{proof}

\begin{Remark}
The above proposition implies, in particular, that the cardinality of the set of scaling limit states is independent of the theory under consideration. Although this may seem surprising at first sight, it must be kept in mind that this doesn't mean that the physical interpretation of these states is the same for all theories: if, for instance, two states of a theory give rise to isomorphic scaling limit nets, in general this will not happen for the corresponding states of another theory. Therefore, upon identifying isomorphic scaling limit theories, we see that the number of the physically distinguishable scaling limits will be different for different theories.
\end{Remark} 

In view of the above result, given a state $\uoo \in \SLF(\o)$, we
will denote by $(\sFiz, U^{(i)}_0,V^{(i)}_0,\Oiz)$, $i=1,2$, the
scaling limit (bosonic) QFTGAs
arising from the corresponding states in
$\textup{SL}^{\sF^{(i)}}(\o^{(i)})$, without further specifications. We will also denote by $\TifpO$ the nuclear maps associated to the theory $\sFi$, $i=1,2$.

\begin{lemma}\label{lem:approxtildeF}
Assume that both theories $\sFi$, $i=1,2$, are asymptotically $p$-nuclear for $p \in (0,1/6)$ and let $\uoo$ be a scaling limit state of $\sF$. For each $\uF \in \ufF(O)$ and each $\eps > 0$, there exists a $\uG \in \tilde{\ufF}$ such that
\begin{equation*}
\norm{\pz(\uF)\Oz - \pz(\uG)\Oz} < \eps.
\end{equation*}
\end{lemma}

\begin{proof}
Without restriction to generality, we can assume $\norm{\uF} \leq 1$. To begin with, we choose $\b > 0$ and non-negative functions $\psi_i \in C(G_i)$, $i=1,2$, which integrate to one, such that, if $\psi(g_1,g_2) := \psi_1(g_1)\psi_2(g_2)$,
\begin{equation*}
\norm{(\Id-e^{-\beta H}\hV_0(\psi))\pz(\uF)\Oz} < \frac{\eps}{2},
\end{equation*}
and we pick $f \in \sS(\bR^d)$ such that $f(p) = e^{-\b p_0}$ for $p \in \Vpc$ (it is straightforward to explicitly construct such a function). Let also $M_i := \limsup_{\l \to 0} \norm{\Th^{(i)}_{\l \b,\l O}}_q$ and let $\d > 0
$ be such that $\d(M_1+M_2+\d) < \eps/2$, and, according to lemma~\ref{lem:approxtheta}, let $\uFi_n \in \ufFi(\hO)$, $n=1,\dots,N$, $i=1,2$, be such that
\begin{equation*}
\limsup_{\k \in K} \bignorm{\TiflkplkO - \PilkN \TiflkplkO}_q \leq \d,
\end{equation*}
for $1\geq q > 2p/(1-4p)$, with obvious meaning of the symbols.
If we define then
\begin{equation*}
\RlN := \sum_{n,m=1}^N \bigscalar{\TflplhO\big(\uFo_{n\l}\otimes\uFt_{m\l}\big)}{\cdot}\TflplhO\big(\uFo_{n\l}\otimes\uFt_{m\l}\big),
\end{equation*}
we have  $\TfpO(F^{(1)}\otimes F^{(2)}) = \TofpO(F^{(1)})\otimes\TtfpO(F^{(2)})$ and $\RlN = \PolN\otimes\PtlN$, and therefore, observing that, by the arguments in~\cite{Mohrdieck:2002a}, the nuclear $q$-norm of $\TflplO - \RlN\TflplO$ agrees with that of its restriction to the the minimal tensor product $\sFo(\l O)\otimes_{\text{min}}\sFt(\l O)$, we can apply lemma~\ref{lem:tensornuclear} in appendix~\ref{app:nuclear}, obtaining 
\begin{equation*}\begin{split}
\bignorm{&\TflplO - \RlN\TflplO}_q \\
&\quad= \bignorm{\ToflplO\otimes\TtflplO - \PolN\ToflplO\otimes\PtlN\TtflplO}_q\\
                           &\quad\leq \bignorm{\ToflplO-\PolN \ToflplO}_q \bignorm{\Th^{(2)}_{\l \b,\l O}}_q\\
&\quad\quad+ \bignorm{\PolN\ToflplO}_q\bignorm{\TtflplO-\PtlN\TtflplO}_q,
\end{split}\end{equation*}
and then
\begin{equation*}
\limsup_\k \bignorm{\TflkplkO-\RlkN\TflkplkO}_q \leq M_2\d +(M_1+\d)\d < \frac{\eps}{2}.
\end{equation*}
We define then the bounded functions
\begin{equation*}
c_{nm}(\l) := \bigscalar{\TflplhO(\uFo_{n\l}\otimes\uFt_{m\l})}{\TflplO(\uF_\l)}, \qquad n,m = 1,\dots,N,
\end{equation*}
and we set $\uH_\l := \sum_{n,m=1}^N c_{nm}(\l)\uFo_{n\l}\otimes \uFt_{m\l}$, $\uH \in \tilde{\ufF}(\hO)$. Since $\tilde{\ufF}$ is a C$^*$-algebra on which translations $\ua$ and gauge transformations $\ub$ act norm continuously, we have $\uG := \ua_{\check{f}}\ub_\psi\uH \in \tilde{\ufF}$, and 
\begin{equation*}\begin{split}
\bignorm{\big[\pz(&\uF)-\pz(\uG)\big]\Oz} \\
     &\leq \bignorm{(\Id-f(P)\hV_0(\psi))\pz(\uF)\Oz} + \bignorm{\TzfpO(\pz(\uF))-\TzfphO(\pz(\uH))}\\
     &\leq \frac{\eps}{2} + \lim_\k \Bignorm{\TflkplkO(\uF_{\l_\k}) - \sum_{n,m=1}^N c_{nm}(\l_\k)\TflkplkhO\big(\uFo_{n\l_\k}\otimes\uFt_{m\l_\k}\big)}\\
&= \frac{\eps}{2} + \lim_\k \bignorm{\big(\TflkplkO-\RlkN\TflkplkO\big)(\uF_{\l_\k})} \\
     &\leq \frac{\eps}{2} + \limsup_\k \bignorm{\TflkplkO-\RlkN\TflkplkO}_q < \eps,
\end{split}\end{equation*}
where in the last inequality we have used the fact that the operator norm is majorized by any nuclear $q$-norm with $0 < q \leq 1$.
\end{proof}

\begin{theorem}\label{thm:tensor}
Assume that the theories $\sFi$, $i=1,2$, are asymptotically $p$-nuclear for $0 < p < 1/6$, and that, for a given scaling limit state $\uoo$ of the tensor product theory $\sF$, the scaling limit theories $\sFiz$, $i=1,2$, satisfy Haag duality. Then there is a unitary equivalence
\begin{equation*}
\sFoz(O) \vntensor \sFtz(O) \ueq \sFz(O)
\end{equation*}
which implements a net-isomorphism between $\sFoz\vntensor\sFtz$ and $\sFz$.
\end{theorem}

\begin{proof}
In view of the last lemma, the vectors $\pz(\uG)\Oz$ with $\uG \in \tilde{\ufF}$ span a dense subspace of the scaling limit Hilbert space $\sHz$. Therefore the operator $W : \sHz \to \sHoz\otimes\sHtz$ defined by
\begin{equation*}
W\pz(\uFo\otimes\uFt)\Oz := \poz(\uFo)\Ooz\otimes\ptz(\uFt)\Otz, \qquad \uFi \in \ufFi, 
\end{equation*}
is unitary, and it is obviously such that
\begin{equation*}
W\pz(\tilde{\ufF}(O))^-W^* = \sFoz(O) \vntensor \sFtz(O).
\end{equation*}
Therefore, identifying unitarily equivalent nets, we get
\begin{equation*}
\sFoz(O) \vntensor \sFtz(O) \subseteq \sFz(O),
\end{equation*}
but $\sFoz\vntensor\sFtz$ satisfies Haag duality, and $\sFz$ satisfies locality by the results of~\cite{D'Antoni:2003ay}, so that, since a net satisfying Haag duality is a maximal local net, the two nets coincide. It is then straightforward to verify that $\ad W$ defines a net-isomorphism from $\sFz$ to $\sFo\vntensor\sFt$.
\end{proof}

\begin{Remarks}
\begin{remlist}
\item According to~\cite[thm.\ 3.4]{Buchholz:1998vu} and to theorem~\ref{thm:free} in the following section, examples in which the nets $\sFiz$ satisfy Haag duality are obtained by taking for $\sFi$, $i=1,2$, nets generated by free fields. In this case, in fact, the corresponding scaling limit theories are also free fields.
\item Another class of examples is obtained, as in the following section, by taking for $\sFo$, say, a net with a classical scaling limit $\sFoz = \bC\Id$, and for $\sFt$ a net with a scaling limit satisfying Haag duality (a free field, for instance). In such cases we have $\sFoz(O) \vntensor \sFtz(O) \ueq \sFtz(O)$, so the scaling limit of the tensor product theory coincides, by theorem~\ref{thm:tensor}, with the scaling limit of the second factor.
\item It is fairly straightforward to modify the proofs of the above results in order to treat the case of Poincar\'e covariant nets, with associated scaling algebras defined by requiring the continuity condition~\eqref{eq:cont} to hold also with respect to Lorentz transformation (these scaling algebras are therefore smaller than those considered above). In particular, under the assumption of asymptotic $p$-nuclearity, the generalized version of lemma~\ref{lem:approxtildeF} will imply that $\sHz = \sHoz\otimes\sHtz$ in this case also. If we assume then geometric modular action for the theories $\sFi$, this will also hold for the scaling limit theories $\sFiz$, $\sFz$~\cite[prop.\ 3.1]{D'Antoni:2003ay}, and this will imply, without further assumptions (in particular without assuming Haag duality in the scaling limit), $\sFz(W) = \sFoz(W)\vntensor\sFtz(W)$ for any wedge $W$, and, as a consequence, equality of the dual theories of $\sFz$ and of $\sFoz\vntensor\sFtz$.
\item Another possible generalization of the results discussed in this section is obtained by dropping the hypothesis that the nets $\sFi$ are purely bosonic. In this case one defines the bosonic and fermionic parts of $\sFi(O)$ as $\sFi(O)_\pm := \{\frac{1}{2}(F\pm\b^{(i)}_{k_i}(F))\,:\,F\in\sFi(O)\}$ and, in order to get a $\bZ_2$-graded theory, the definition of the tensor product theory given above must be altered by replacing equation~\eqref{eq:tensorF} with
\begin{equation*}
\sF(O) := \sFo(O)\hat{\otimes}\sFt(O):=\sFo(O)\vntensor\sFt(O)_+ + \Vo(k_1)\sFo(O)\vntensor\sFt(O)_-.
\end{equation*} 
The analysis proceeds then along the same lines as the above one, by studying the $\l \to 0$ behaviour of the nuclearity properties of the restrictions of the maps $\TflplO$ to the bosonic and fermionic subnets, and at the end one obtains that if the theories $\sFi$ are asymptotically $p$-nuclear and their scaling limit theories $\sFiz$ satisfy twisted Haag duality, then $\sFz$ and $\sFoz\hat{\otimes}\sFtz$ are net-isomorphic.
\end{remlist}
\end{Remarks}

We will need a version of theorem~\ref{thm:tensor} which deals with outer regularized scaling limit nets:
\begin{equation*}
\sF_{0,r}(O) := \bigcap_{O_1 \supset \overline{O}} \pz(\ufF(O_1))''.
\end{equation*}

\begin{theorem}\label{thm:tensorreg}
Assume that the purely bosonic theories $\sFi$, $i=1,\dots,n$, are asymptotically $p$-nuclear for $0 < p < 1/6$, and that the outer regularized scaling limit theories $\sF^{(i)}_{0,r}$, $i=1,\dots,n$, satisfy Haag duality. Then
\begin{equation*}
\sF^{(1)}_{0,r}(O) \vntensor\dots\vntensor \sF^{(n)}_{0,r}(O) \ueq \sF_{0,r}(O)
\end{equation*}
and the theories $\sF^{(1)}_{0,r}\vntensor\dots\vntensor\sF^{(n)}_{0,r}$ and $\sF_{0,r}$ are net-isomorphic.
\end{theorem}

\begin{proof}
According to lemma~\ref{lem:tensornuclear}, the tensor product of two asymptotically $p$-nuclear theories is again asymptotically $p$-nuclear, and therefore it is sufficient, by induction, to prove the theorem for $n=2$.
 We begin by observing that, if $(M_\a)_\a$, $(N_\a)_\a$ are families of von Neumann algebras (on the same Hilbert space), then 
\begin{equation*}
\bigwedge_{\a} M_\a \vntensor N_\a = \Big(\bigwedge_\a M_\a\Big)\vntensor\Big(\bigwedge_\a N_\a\Big).
\end{equation*}
Through the unitary equivalence induced by the operator $W : \sHz\to \sHoz\otimes\sHtz$ defined in the proof of theorem~\ref{thm:tensor}, we have
\begin{equation*}
\poz(\ufFo(O))\otimes_{\text{min}}\ptz(\ufFt(O)) \subseteq \pz(\ufF(O)),
\end{equation*}
and then, by what we have just observed,
\begin{equation*}\begin{split}
\sF^{(1)}_{0,r}(O) \vntensor \sF^{(2)}_{0,r}(O) &= \Big(\bigcap_{O_1 \supset \overline{O}} \poz\big(\ufFo(O_1)\big)''\Big)\vntensor\Big(\bigcap_{O_1 \supset \overline{O}} \ptz\big(\ufFt(O_1)\big)''\Big) \\
&= \bigcap_{O_1 \supset \overline{O}}\poz\big(\ufFo(O_1)\big)''\vntensor\ptz\big(\ufFt(O_1)\big)''\subseteq \bigcap_{O_1 \supset \overline{O}} \pz\big(\ufF(O_1)\big)'' \\ &= \sF_{0,r}(O),
\end{split}\end{equation*}
and we conclude, as in the proof of theorem~\ref{thm:tensor}, using the maximality of Haag dual nets.
\end{proof}

\section{A class of models with non-preserved DHR sectors}\label{sec:model}
In this section we will construct quantum field theory models which possess DHR sectors which are non-preserved in the scaling limit. As already said in the introduction, the observables net of such models is obtained as the fixed point net of a field net which in turn is a tensor product of two theories, one being a theory with trivial scaling limit, and the other being a free field theory, whose DHR sectors are all preserved. We start then by briefly recalling some facts about theories with trivial scaling limit and the scaling limit of free field theories from~\cite{Lutz:1997a, Buchholz:1998vu}.

Let $\phi$ be the generalized free scalar field with mass measure $d\r(m) = dm$ in $d=s+1=3,4$ spacetime dimensions, i.e.
\begin{equation}\label{eq:generalfree}
\phi(f) = \frac{i}{\sqrt{2}}[a(T\bar{f})-a(Tf)^*], \qquad f \in \sS(\bR^d),
\end{equation}
with $a(\psi)$, $a(\psi)^*$ being the annihilation and creation operators on the symmetric Fock space over $L^2(\bR^s\times\bR_+,d^s\bp \,dm)$ and 
\begin{equation*}
Tf(\bp,m) := (2\o_m(\bp))^{-\frac{1}{2}}\hat{f}(\o_m(\bp),\bp),\qquad \o_m(\bp) := \sqrt{m^2+\abs{\bp}^2}.
\end{equation*}
Let also $\l \in \bR_+ \to n(\l) \in \bN_0$ be a non-increasing function which diverges as $\l \to 0$, and for a double cone $O = (x+\Vp)\cap(y-\Vp)$ define $n(O) := n(\sqrt{(x-y)^2})$, and consider the net of local algebras
\begin{equation}\label{eq:lutz}
\sAL(O) := \{e^{i\phi(f)}\,:\, f\in \Box^{n(O)}\sD_\bR(O)\}'',
\end{equation}
with the obvious definition of the action of translations. In~\cite{Lutz:1997a}, developing an idea exposed in~\cite{Buchholz:1996mx}, it is shown that the net $\sAL$ satisfies the standard assumptions, including weak additivity and essential Haag duality, and that the corresponding operators $\Th^L_{\b,O}$, defined as in equation~\eqref{eq:nuclearmap}, satisfy
\begin{equation}\label{eq:lutznuclear}
\limsup_{\l \to 0}\norm{\Th^L_{\l\b,\l O}}_p \leq 1,
\end{equation}
for all $\beta$, $O$ and all $0 < p \leq 1$, and therefore, according to proposition~\ref{prop:nuclearlimit}.\ref{part:classical}, the net $\sAL$ has classical scaling limit.

We now consider the scaling limit of the free scalar field. Following~\cite{Buchholz:1998vu}, we use a non-standard, locally Fock representation of the local algebras in the Cauchy-data formulation of the free field. Denote by $\tfW$, the (abstract) Weyl algebra over $(\sD(\bR^s),\ts)$, where 
\begin{equation*}
\ts(f,g) = \Im \int_{\bR^s}d^s\bx\overline{f(\bx)}g(\bx),
\end{equation*}
equipped with (mass dependent) actions $\amLx$ of the Poincar\'e group and $\td_\l$ of dilations, and let $\om$ denote the vacuum state with mass $m$ (see~\cite{Buchholz:1998vu} for the explicit formulas, which will not be needed in the following).

The non-standard free field representation used in~\cite{Buchholz:1998vu} is obtained in the following way. For an open ball $B \subset \bR^s$, denote by $\tfW(B)$ the Weyl algebra over $(\sD(B),\ts)$. Then the states $\om \rest \tfW(B)$ and $\oz \rest\tfW(B)$ are known to be normal to each other~\cite{Eckmann:1974a}, and then, if $\pim$, $\piz$ are the GNS representations of $\tfW$ induced by $\om$, $\oz$, the von Neumann algebras $\pim(\tfW(B))''$ and $\piz(\tfW(B))''$ are isomorphic through an isomorphism connecting $\pim(W)$ and $\piz(W)$. It is also straightforward to verify that, using these isomorphisms, the action $\am$ of $\Pport$ on $\tfW$ can be transported to an action on $\piz(\tfW)$, still denoted by $\am$. Therefore, by defining,
\begin{equation*}
\sAm(\L O_B + x) := \big\{\amLx\big(\piz(\Wt(g))\big)\,:\,g \in \sD(B)\big\}'',
\end{equation*}
with $O_B = B''$ the causal completion of the $s$-dimensional ball $B$, one gets a net of local von Neumann algebras for the scalar field of mass $m$ which is represented on the Fock space $\sHuo$ of the field of mass zero, but which is isomorphic to the mass $m$ net in the usual Fock representation. We also note that, $\oz$ being $\td_\l$-invariant, $\td_\l$ is unitarily implemented in this representation (but of course it is not an automorphism of the net $\sAm$, unless $m=0$). From now on we will drop the indication of the representation $\piz$.

Let $\fAm(O)$ be the C$^*$-algebra of the elements $A \in \sAm(O)$ such that $x \to \am_x(A)$ is norm-continuous. Due to the outer continuity of the net $\sAm$, $\sAm(O) = \bigcap_{O_1 \supset \overline{O}}\sAm(O_1)$~\cite{Araki:1963a}, we have $\fAm(O)^- = \sAm(O)$. Note that $\fAm$ is also outer continuous. Let then $\uAm$ be the net of scaling algebras associated to $\fAm$, and, with $\uoo$ any scaling limit state, and $\pz$ the corresponding GNS representation, we define $\fAmz$ as the outer regularized scaling limit net
\begin{equation}
\fAmz(O) := \bigcap_{O_1 \supset \overline{O}}\pz(\uAm(O_1)).
\end{equation}

One of the main results of~\cite{Buchholz:1998vu} is that the formula
\begin{equation}\label{eq:phitilde}
\tth_m(\pz(\uA)) = w\text{-}\lim_\k\td_{\l_\k}^{-1}(\uA_{\l_\k}),
\end{equation}
with $(\l_\k)_\k \subset \bR_+$ a net such that $\uoo = \lim_\k \uo_{\l_\k}$, defines a net isomorphism between $\fAmz$ and $\fA^{(0)}$, which is implemented by a unitary $\tV_m : \sHz \to \sHuo$. 

\begin{proposition}\label{prop:isoreg}
With the above notations, $\tth_m = \ad \tV_m$ extends to a net isomorphism between the outer regularized net of von Neumann algebras
\begin{equation}
\sAmz(O) := \bigcap_{O_1 \supset \overline{O}}\pz(\uAm(O_1))'',
\end{equation}
and $\sAz$.
\end{proposition}

\begin{proof} 
We have, by the outer regularity of $\sAz$,
\begin{equation*}
\tV_m \fAmz(O)'' {\tV}^*_m = \sAz(O) = \bigcap_{O_1 \supset \overline{O}} \sAz(O_1) = \tV_m\bigg(\bigcap_{O_1 \supset \overline{O}} \fAmz(O_1)''\bigg){\tV}^*_m,
\end{equation*}
and therefore
\begin{equation*}\begin{split}
\bigcap_{O_1 \supset \overline{O}} \fAmz(O_1)'' &= \bigg(\bigcap_{O_1 \supset \overline{O}} \pz(\uAm(O_1))\bigg)'' \\
                                                &\subseteq \bigcap_{O_1 \supset \overline{O}} \pz(\uAm(O_1))''\subseteq \bigcap_{O_1 \supset \overline{O}} \fAmz(O_1)'',
\end{split}\end{equation*}
so that, finally,
\begin{equation*}
\tth_m(\sAmz(O)) = \tth_m\bigg( \bigcap_{O_1 \supset \overline{O}} \pz(\uAm(O_1))\bigg)'' = \sAz(O).
\end{equation*}
\end{proof}

We now pass to the construction of the class of models mentioned above. Let $G$ be a compact Lie group and $N \subset G$ a (proper and nontrivial) normal closed subgroup. It is then possible~\cite[thm.\ 6.1.1]{Price:1977a} to find a finite set $\D$ of irreducible representations of $G$ which is \emph{symmetric} and \emph{generating}, i.e.\ such that for each representation belonging to $\D$ also its conjugate representation is in $\D$, and every irreducible representation of $G$ is a subrepresentation of a tensor product of representations from $\D$. Denote now by $\D_2$ the subset of $\D$ of those representations which are trivial on $N_2 := N$, and define another closed normal subgroup $N_1$ of $G$ as the annihilator of $\D_1 := \D\setminus\D_2$: $N_1 := \bigcap_{v \in \D_1} \ker v$. It follows then easily that $G$ is isomorphic to $G_1\times G_2$, with $G_i := G/N_i$, $i=1,2$, and that $\D_i$ is a symmetric generating set of irreducible representations for $G_i$. We introduce the representations $v_i := \bigoplus_{v\in\D_i}v$ of $G_i$, and set $n_i := \dim v_i$, $i=1,2$. Since $\D_i$ is symmetric, there exists a unitary involution $J_i$ on $\bC^{n_i}$ such that $J_iv_i(\cdot)J_i$ is the complex conjugate representation of $v_i$, and it is possible to find $m_i, p_i \in \bN$ with $n_i = 2m_i+p_i$, and vectors $(e_k^{(i)})_{k=1,\dots,m_i+p_i}$ of $\bC^{n_i}$ such that $\{e^{(i)}_1,\dots,e^{(i)}_{m_i+p_i},J_ie^{(i)}_1,\dots,J_ie^{(i)}_{m_i}\}$ is an orthonormal basis and $J_i e^{(i)}_k = e^{(i)}_k$ for $k=m_i+1,\dots,m_i+p_i$ (i.e.\ $e^{(i)}_{m_i+1},\dots,e^{(i)}_{m_i+p_i}$ span the subspace of the real representations in $\D_i$).

The first factor $\sFo$ of our class of models is defined as the field net generated by a $v$-multiplet of generalized free fields~\eqref{eq:generalfree} for each $v\in\D_1$, to which a suitable power of the Dalembertian has been applied, as in~\eqref{eq:lutz}. More specifically, on the symmetric Fock space $\sHo$ over $L^2(\bR^s\times\bR_+,d^s\bp \,dm)\otimes\bC^{n_1}$, with vacuum vector $\Oo$, we consider the fields
\begin{equation*}
\phi_k(f) = \frac{i}{\sqrt{2}}[a(T\bar{f}\otimes e^{(1)}_k)-a(Tf\otimes J_1e^{(1)}_k)^*], \quad f \in \sS(\bR^d),\, k=1,\dots,m_1+p_1,
\end{equation*}
($\phi_1,\dots ,\phi_{m_1}$ are complex generalized free fields, while $\phi_{m_1+1},\dots, \phi_{m_1+p_1}$ are real generalized free fields), and, assuming that there is a $\l_0$ such that $n(\l) = 0$ for $\l\geq \l_0$, define the net of von Neumann algebras
\begin{equation*}
\sFo(O) := \{e^{i[\phi_k(f)+\phi_k(f)^*]^-}\,:\, f\in \Box^{n(O)}\sD_\bR(O),\,k=1,\dots,m_1+p_1\}'',
\end{equation*}
with the obvious action $\ao_x := \ad \Uo(x)$ of the translations and with the action $\bo_g := \ad \Vo(g)$ of $G_1$, where $\Vo(g)$ is the second quantization of $\Id\otimes v_1(g)$.

\begin{proposition}\label{prop:tensorlutz}
The bosonic theory $(\sFo,\Uo,\Vo,\Oo)$ is asymptotically $p$-nuclear for each $p \in (0,1]$ and has a classical scaling limit. The associated net of observables $\sAo := (\sFo)^{G_1}$ has DHR sectors in one to one correspondence with the unitary equivalence classes of irreducible representations of $G_1$.
\end{proposition}

\begin{proof}
It is a straightforward consequence of the fact that $\sFo \ueq \sA_L^{\otimes n_1}$ and of the by now common argument involving lemma~\ref{lem:tensornuclear}, that for the nuclear operator $\Th^{(1)}_{\b,O}$ of $\sFo$ there holds $\Th^{(1)}_{\b,O} = (\Th^L_{\b,O})^{\otimes n_1}$ and $\norm{\Th^{(1)}_{\b,O}}_p\leq \norm{\Th^L_{\b,O}}_p^{n_1}$ for each $p\in(0,1]$, and therefore it follows from equation~\eqref{eq:lutznuclear} that $\sFo$ is asymptotically $p$-nuclear and from proposition~\ref{prop:nuclearlimit}.\ref{part:classical} that it has classical scaling limit.

For what concerns the second part of the statement, it follows easily from the results in~\cite{Lutz:1997a} that the theory under consideration satisfies the assumptions 1.-7. in~\cite{Doplicher:1969tk}, and this entails that the DHR sectors of $\sAo$ which appear in $\sHo$ are in one to one correspondence with the classes of irreducible representations of $G_1$~\cite[thm. 3.6]{Doplicher:1969tk}.
\end{proof}

\begin{Remark} It may seem natural to conjecture that the net $\sAo$ has no other DHR sector apart from those described by the group $G_1$. The standard way to prove this, would be to show that the net $\sFo$ satisfies the split property and a certain cohomological condition~\cite[sec.\ 3.4.5]{Roberts:1989ps}. While it is straightforward to verify that this is actually the case, as $\sAo$ does not satisfy Haag duality this can only be used to conclude that classes of irreducible representations of $G_1$ label classes of local 1-cocycles of $\sAo$~\cite{Roberts:1989ps}. On the other hand, it follows from standard arguments that the dual net $\smash{\sAo}^d$ coincides with the $G_1$-fixed point net of the dual of the field net generated by the generalized free fields $\phi_k(f)$ (without further conditions on the test functions $f$) and such a net doesn't even satisfy the split property~\cite{Doplicher:1984a}. It is therefore an open problem to determine the complete superselection structure of $\sAo$.
\end{Remark}

The factor $\sFt$ in our model is the net generated by multiplets of free scalar fields belonging to the representations in $\D_2$. We consider therefore the Weyl algebra $\fW$ over $(\sD(\bR^s)\otimes \bC^{n_2},\s)$ with
\begin{equation*}
\s(f\otimes\x,g\otimes\e) = \Im \bigg[\int_{\bR^s}d^s\bx\overline{f(\bx)}g(\bx)(\x\cdot\e)\bigg],
\end{equation*}
where $(\x\cdot\e)$ is the standard scalar product of $\x,\e\in\bC^{n_2}$, and pick a mass function $\bm : \D_2 \to \overline{\bR}_+$ such that $\bm(v) = \bm(\bar{v})$. As $\fW$ is isomorphic with $\tfW^{\otimes_{\text{min}}n_2}$, we define an action $\abm$ of $\rPport$, an action $\d$ of dilations and a vacuum state $\obm$ through
\begin{align*}
\abm_\Lx &:= \a^{(\m_1)}_\Lx \otimes \dots \otimes \a^{(\m_{n_2})}_\Lx, \\
\d_\l &:= \td^{\otimes n_2}_\l, \\
\obm &:= \o^{(\m_1)}\otimes\dots\otimes\o^{(\m_{n_2})},
\end{align*}
where $(\m_1,\dots,\m_{n_2})$ is a vector obtained by repeating each value $\bm(v)$ exactly $\dim v$ times, for each $v\in\D_2$. Furthermore, an action $\bt$ of $G_2$ is defined by
\begin{equation*}
\bt_g(W(f)) := W\big((\Id\otimes v_2(g))f\big), \qquad f \in \sD(\bR^s)\otimes\bC^{n_2}, g \in G_2.
\end{equation*}
If $\bze: \D_2 \to \overline{\bR}_+$ is the identically vanishing function, it is easy to verify that the states $\obm$ and $\obz$ are locally normal to each other, and therefore, in analogy to the $n_2 =1$ case, we define, in the GNS representation of $\obz$, the net
\begin{equation*}
\sFbm(\L O_B +x) := \Big\{\abm_\Lx\big(W(f)\big)\,:\, f \in \sD(B)\otimes\bC^{n_2}\Big\}''.
\end{equation*}
where we suppressed the explicit indication of the GNS representation. The corresponding unitary representations of $\rPport$ and $G_2$ will be denoted by $\Ubm$, $\Vt$, and the vacuum vector by $\Obm$

Consider now a scaling limit state $\uobmo$ of the scaling algebra $\uFbm$ associated to $\sFbm$, and the corresponding GNS representation $\pz$. In analogy to the scalar field case, we define the corresponding (outer regularized) scaling limit net $\sFbmz$ by
\begin{equation}
\sFbmz(O) := \bigcap_{O_1 \supset \overline{O}} \pz(\uFbm(O_1))''.
\end{equation}

\begin{theorem}\label{thm:free}
Let $s=2,3$, and $\uobmo$ be a scaling limit state of $\uFbm$. There is a net-isomorphism $\th$ between $(\sFbmz, \Ubmz, \Vtz, \Obm_0)$ and $(\sFbz, \Ubz, \Vt, \Obz)$, which is unitarily implemented and is such that
\begin{equation}
\th(\pz(\uF)) = w\text{-}\lim_\k\d_{\l_\k}^{-1}(\uF_{\l_\k}),
\end{equation}
with $(\l_\k)_\k \subset \bR_+$ a net such that $\uobmo = \lim_\k \uo^{(\bm)}_{\l_\k}$.
\end{theorem}

\begin{proof}
We begin by introducing an auxiliary scaling algebra $\uGbm$, which is defined as the scaling algebra associated to the net $(\sFbm, \Ubm,\Obm)$, i.e. which disregards the action of the gauge group. Of course $\uFbm(O) \subseteq \uGbm(O)$ for each $O$, and there exists a scaling limit state of $\uGbm$ whose restriction to $\uFbm$ coincides with $\uobmo$. By a slight abuse of notation, we denote by $\pz$ the scaling limit representation of $\uGbm$ thus obtained, and we define
\begin{equation}
\sGbmz(O) := \bigcap_{O_1 \supset \overline{O}} \pz(\uGbm(O_1))''.
\end{equation}

From the net isomorphism
\begin{equation*}
(\sFbm, \Ubm, \Obm) \iso (\sA^{(\m_1)}\vntensor\dots\vntensor\sA^{(\m_{n_2})},U^{(\m_1)}\otimes\dots\otimes U^{(\m_{n_2})},\O^{(\m_1)}\otimes\dots\otimes\O^{(\m_{n_2})}),
\end{equation*}
taking into account the fact that, according to~\cite[prop.\ 3.1]{Mohrdieck:2002a}, each theory $\sA^{(\m_k)}$ is asymptotically $p$-nuclear for each $p \in (0,1]$, and the fact that, according to proposition~\ref{prop:isoreg}, each scaling limit net $\sA_0^{(\m_k)}$ satisfies Haag duality, it follows, applying theorem~\ref{thm:tensorreg}, that we have a net isomorphism $\th$
\begin{equation*}\begin{split}
(\sGbmz, \Ubmz, \Obm_0) &\iso (\sA^{(0)}\vntensor\dots\vntensor\sA^{(0)},U^{(0)}\otimes\dots\otimes U^{(0)},\O^{(0)}\otimes\dots\otimes\O^{(0)})\\
&\iso (\sFbz, \Ubz, \Obz),
\end{split}\end{equation*}
such that, for each $\uG$ of the form  $\uG_\l = \uA_{1\,\l}\otimes \dots \otimes \uA_{n_2\,\l}$, $\uA_k \in \ufA^{(\m_k)}$, 
\begin{equation}\begin{split}\label{eq:phi}
\th(\pz(\uG)) &= (\tth_{\m_1} \otimes \dots \otimes \tth_{\m_{2n}})\big((\pz^{(\m_1)}\otimes\dots\otimes\pz^{(\m_{2n})})(\uG)\big) \\
               &= w\text{-}\lim_\k(\td_{\l_\k}^{-1}\otimes\dots\otimes\td_{\l_\k}^{-1})(\uG_{\l_\k}) \\
               &= w\text{-}\lim_\k\d_{\l_\k}^{-1}(\uG_{\l_\k}),
\end{split}\end{equation}
where $\pz^{(\m_k)}$ is the scaling limit representation of $\ufA^{(\m_k)}$ induced by the scaling limit state $\uA_k \to \uobmo(\underline{\Id}\otimes\dots\otimes \uA_k\otimes \dots\otimes\underline{\Id})=\uoo^{(\m_k)}(\uA_k)$. The last relation is then extended by linearity and continuity to the C$^*$-subalgebra $\tuGbm$ of $\uGbm$ generated by all such functions. Furthermore $\th = \ad (\tV_{\m_1}\otimes\dots\otimes\tV_{\m_{2n}})$. We now extend equation~\eqref{eq:phi} to arbitrary  elements of $\uGbm$. From the proof of theorem~\ref{thm:tensorreg} it follows that $\pz(\uGbm(O))\subset \bigcap_{O_1\supset\overline{O}} \pz(\tuGbm(O_1))''$, and, therefore, for $\uF \in \uGbm(O)$ and for each $O_B \supset \overline{O}$ and $\eps > 0$, we can find $\uG \in \tuGbm(O_B)$ such that 
\begin{equation}\label{eq:approxF}
\norm{[\pz(\uF)-\pz(\uG)]\Oz} <\eps.
\end{equation}
Furthermore, if $\uH \in \uGbm(O_B')$, we have
\begin{equation}\label{eq:phiF}\begin{split}
\bigscalar{\th(\pz(\uH))\O^{(\bze)}}{&\big[\th(\pz(\uF))-\d_{\l_\k}^{-1}(\uF_{\l_\k})\big]\th(\pz(\uH))\O^{(\bze)}} \\
                           &= \bigscalar{\pz(\uH^*\uH)\Oz}{\big[\pz(\uF)-\pz(\uG)]\Oz}\\
                           &\quad + \bigscalar{\th(\pz(\uH))\O^{(\bze)}}{\big[\th(\pz(\uG))-\d_{\l_\k}^{-1}(\uG_{\l_\k})\big]\th(\pz(\uH))\O^{(\bze)}}\\
                           &\quad + \bigscalar{\d_{\l_\k}\big(\th(\pz(\uH^*\uH))\big)\O^{(\bze)}}{\big(\uF_{\l_\k}-\uG_{\l_\k}\big)\O^{(\bze)}},
\end{split}\end{equation}
and the three terms of the right hand side of such equation can be made arbitrarily small, for sufficiently large $\k$, thanks, respectively, to~\eqref{eq:approxF}, to~\eqref{eq:phi} and to the fact that, since $\obm$ and $\obz$ are locally normal to each other, we have, by~\eqref{eq:roberts},
\begin{equation*}\begin{split}
\lim_\k \norm{(\uF_{\l_\k}&-\uG_{\l_\k})\O^{(\bze)}}^2 = \lim_\k\obz\big((\uF_{\l_\k}-\uG_{\l_\k})^*(\uF_{\l_\k}-\uG_{\l_\k})\big) \\
&= \lim_\k \obm\big((\uF_{\l_\k}-\uG_{\l_\k})^*(\uF_{\l_\k}-\uG_{\l_\k})\big) = \norm{[\pz(\uF)-\pz(\uG)]\Oz}^2.
\end{split}\end{equation*}
Then, since $\th(\pz(\uGbm(O_B')))\O^{(\bze)}$ is dense in $\sH^{(\bze)}$ (its closure contains $\sFbz(O_1')\O^{(\bze)}$ for each $O_1 \supset O_B$) and $(\d_{\l_\k}^{-1}(\uF_{\l_\k}))_\k$ is a bounded net, we conclude that~\eqref{eq:phi} holds for each element of $\uGbm$.

In order to complete the proof, we need only show that $\sGbmz(O) = \sFbmz(O)$. To this end, let $\uF \in \uGbm(O)$, and, for $\psi \in L^1(G)$, consider
$\ubt_\psi \uF \in \uFbm(O)$. Thanks to~\eqref{eq:phi} and to the fact that $\d_\l$ and $\bt_g$ are commuting and unitarily implemented on $\sH^{(\bze)}$, we have that, for each $\Phi \in \sH^{(\bze)}$, we can find a sequence $(\l_n)_{n \in \bN}$ converging to 0 such that
\begin{equation*}
\scalar{\Phi}{\th(\pz(\ubt_\psi\uF))\Phi} = \lim_{n \to +\infty} \int_G dg\,\psi(g)\scalar{\Phi}{\bt_g\d_{\l_n}(\uF_{\l_n})\Phi}.
\end{equation*}
But, using again~\eqref{eq:phi}, we have $\lim_{n \to +\infty}\scalar{\Phi}{\bt_g\d_{\l_n}(\uF_{\l_n})\Phi} =\scalar{\Phi}{\bt_g(\th(\pz(\uF)))\Phi}$, and therefore, applying the dominated convergence theorem,
\begin{equation*}
\th(\pz(\ubt_\psi\uF)) = \int_G dg\,\psi(g)\bt_g(\th(\pz(\uF))).
\end{equation*}
This last equation entails that, if $(\psi_n)_{n \in \bN}$ is a $\d$-sequence in $L^1(G)$, then $\th(\pz(\ubt_{\psi_n}\uF))$ converges strongly to $\th(\pz(\uF))$, i.e. $\th(\pz(\uGbm(O))) \subseteq \th(\pz(\uFbm(O)))'' \subseteq \th(\pz(\uGbm(O)))''$, and the cyclic Hilbert spaces $\overline{\pz(\uGbm)\Oz}$ and $\overline{\pz(\uFbm)\Oz}$ coincide (recall that $\th$ is unitarily implemented), from which the equality of the nets $\sFbmz$ and $\sGbmz$ readily follows.
\end{proof}

Knowing explicitly the scaling limit of $\sFbm$, it is not difficult to show that the superselection structure of the observable net $\sAbm(O) = \sFbm(O)^{G_2}$ is entirely preserved.

\begin{theorem}\label{thm:preserv}
Each covariant, finite statistics sector of the net $\sAbm$ is preserved in each scaling limit state $\uobmo$.
\end{theorem}

\begin{proof}
As recalled in section~\ref{sec:scalingfields}, we must find, for each covariant, finite statistics sector $\xi$ of $\sAbm$ and for each double cone $O$, a scaled multiplet $\psi_j(\l) \in \sFbm(\l O)$, $j=1,\dots, d$, $d$ the statistical dimension of $\xi$, and for each $j$ an $\uF \in \uFbm(O_1)$ with $O_1 \supset \bar{O}$, for which equation~\eqref{eq:preserv} holds.

To this end, let $O = \L O_B + x$, take a multiplet $\psi_j \in \sFbm(O_B) = \sFbz(O_B)$ associated to the sector $\xi$, and define $\psi_j(\l) := \abm_{(\L,\l x)}\d_\l(\psi_j) \in \sFbm(\l O)$. We claim that for this multiplet, equation~\eqref{eq:preserv} is satisfied.

Since $\bt_g$ commutes with $\abm_{(\L,\l x)}\d_\l$, it is obvious that $\psi_j(\l)$, $j=1,\dots,d$, is a multiplet of class $\xi$. Now, since $\th^{-1}(\psi_j) \in \sFbmz(O_B)= \bigcap_{O_2 \supset \overline{O}_B} \pz(\uFbm(O_2))''$, for each $\eps > 0$ and $O_2 \supset \overline{O}_B$, we can find $\uG \in \uFbm(O_2)$ such that
\begin{equation*}
\bignorm{[\th^{-1}(\psi_j)-\pz(\uG)]\Oz} +\bignorm{[\th^{-1}(\psi_j)-\pz(\uG)]^*\Oz}<\eps.
\end{equation*}
We have then the following chain of equalities, where $[\dots]^\sharp$ stands for either $[\dots]$ or $[\dots]^*$:
\begin{equation*}\begin{split}
\bignorm{[\th^{-1}(\psi_j)-\pz(\uG)]^\sharp\Oz} &= \bignorm{[\psi_j-\th(\pz(\uG))]^\sharp\Obz} \\
                                          &= \lim_\k \bignorm{[\psi_j-\d_{\l_\k}^{-1}(\uG_{\l_\k})]^\sharp\Obz}\\
                                          &= \lim_\k \bignorm{[\d_{\l_\k}(\psi_j)-\uG_{\l_\k}]^\sharp\Obz}\\
                                          &= \lim_\k \bignorm{[\d_{\l_\k}(\psi_j)-\uG_{\l_\k}]^\sharp\Obm}\\
                                          &= \lim_\k \bignorm{\big[\psi_j(\l_\k)-\big(\uabm_{(\L,x)}\uG\big)_{\l_\k}\big]^\sharp\Obm}
\end{split}\end{equation*}
 where in the fourth equality we have again applied equation~\eqref{eq:roberts} to $\obm$ and $\obz$.
Therefore, taking into account the remark about Poincar\'e covariance in section~\ref{sec:scalingfields}, we get~\eqref{eq:preserv} with $\uF := \uabm_{(\L,x)}\uG$.
\end{proof}

From the above theorem and from~\cite[cor.\ 6.2]{D'Antoni:2003ay} the following result readily follows.

\begin{corollary}\label{cor:equivalence}
All local intertwiners between DHR endomorphisms of $\sAbm$ are also global intertwiners.
\end{corollary}

We remark that a property closely related to the equivalence of local and global intertwiners has been recently proven, under quite general assumptions, in the context of locally covariant theories~\cite{Brunetti:2005a}. 

The above results may be summarized in the following theorem, which expresses the existence of a rather vast class of decent quantum field theory models possessing non-preserved DHR sectors.

\begin{theorem}\label{thm:main}
For each pair $(G,N)$ with $G$ a compact Lie group and $N \subset G$ a normal closed subgroup, there exists a bosonic QFTGA $(\sF,U,V,\O)$ such that the associated observable net $\sA := \sF^G$ fulfills the following properties:
\begin{proplist}{3}
\item $\sA$ has a subset of DHR sectors which are in 1-1 correspondence with the unitary equivalence classes of irreducible representations of $G$;
\item $\sA$ has a unique quantum scaling limit according to the classification of~\cite{Buchholz:1995a};
\item among the sectors of $\sA$, only those which correspond to representations of $G$ which are trivial on $N$ are preserved in any scaling limit state;
\item the set of DHR sectors of each (outer regularized) scaling limit net $\sA_0$ of $\sA$ is in 1-1 correspondence with the unitary equivalence classes of irreducible representations of $G/N$.
\end{proplist}
\end{theorem}

\begin{proof}
For a given choice of a finite symmetric generating set $\D = \D_1 \cup \D_2$ of irreducible representations of $G$ and a mass function $\bm : \D_2 \to \bR_+$ as above, we form the tensor product theory $\sF := \sFo\otimes\sFbm$. It is easy to check, as in the proof of proposition~\ref{prop:tensorlutz}, that the sectors of $\sA = \sF^G$ which appear in the Hilbert space $\sH$ on which $\sF$ acts are in 1-1 correspondence with classes of irreducibile representations of $G$, thereby proving (i). It then follows from theorem~\ref{thm:tensorreg}, proposition~\ref{prop:tensorlutz} and theorem~\ref{thm:free} that each outer regularized scaling limit net $\sF_0$ of $\sF$ is unitarily equivalent to the net $\sFbz$ and then, since, according to the results in~\cite{Conti:2004a}, to each scaling limit state $\uoo$ of $\sA$ there corresponds uniquely a scaling limit state of $\sF$ whose restriction to $\ufA$ coincides with $\uoo$, statement (ii) follows. Property (iii) is the content of theorem~\ref{thm:preserv}, and finally property (iv) follows from the fact that $\sFbz$, being a finite tensor product of free scalar field nets, satisfies the split property and Roberts' cohomological condition, and it is therefore a complete field net, i.e. all DHR sectors of $\sA_0 \ueq \smash{\sFbz}^{G/N}$ are implemented by $G/N$-multiplets in $\sFbz$.
\end{proof}

As noted in the remark following proposition~\ref{prop:tensorlutz}, it may also happen that $\sA$ has more DHR sectors than those described by $\sF$, but, since $\sA_0$ has precisely the sectors described by $\sFbz$, also these additional sectors would not be preserved under the scaling limit.

\section{Conclusions and outlook}
In this work, we have presented very simple quantum field theory models possessing DHR superselection sectors which are not preserved under the scaling limit operation, i.e. sectors of the underlying theory which are not also sectors of the scaling limit. The way in which these sectors are obtained provides a simple illustration of the physical mechanism which may be expected to lead to the appearance of non-preserved sectors in more realistic, interacting theories. As already discussed in~\cite{D'Antoni:2004a}, charges will disappear in the scaling limit if they have some kind of ``internal structure'' which, in order for it to be ``squeezed'' in a region of radius $\l$, requires an amount of energy growing faster than $\l^{-1}$. Therefore one can expect that the fields carrying such a charge will have rather bad ultraviolet properties, and this is actually the case for the fields $\Box^{n(\l)}\phi_k(x)$ employed in constructing our models at scale $\l$.

As we mentioned above, our examples, since they are built making use of generalized free fields with constant mass measure, do not satisfy Haag duality, but only essential duality. This leaves open the possibility that requiring Haag duality rules out the existence of non-preserved sectors, but, in view of the physical picture just discussed, one may be tempted to exclude that this is actually the case.

As a tool for building the models, we have derived sufficient conditions under which the scaling limit of a tensor product theory coincides with the tensor product of the scaling limits of the factor theories, the main such condition being a requirement of asymptotic nuclearity for the factor theories. While we don't have any example of theories not satisfying such hypothesis for which the scaling limit and tensor product operations don't commute, it seems to us quite natural that some kind of phase space condition has to play a role in such questions, particularly in view of the fact that, if a specific such condition holds, scaling limits are limits with respect to a suitable metric in a suitable space of nets of C$^*$-algebras~\cite{Guido:2005a}, and therefore they should enjoy good functorial properties. 

This is connected with the fact that we required $G$ to be not just a compact group, but a Lie one, which is due to the following technical reason: according to the results in~\cite{Mohrdieck:2002a}, an infinite tensor product of free scalar field theories is not asymptotically $p$-nuclear for $0 < p<1/3$, and therefore we have to use, in our construction, pairs $(G,N)$ for which the set $\D_2$ which generates the representations which are trivial on $N$ is finite, i.e., even if we just assume that $G$ is a compact group, we have in any case to require $G/N$ to be a Lie group. It is therefore not clear if it is possible to construct, along the lines exposed above, examples of theories having, as in~\cite{Doplicher:2002cb}, an arbitrary compact group $G$ as a gauge group, and such that the sectors associated to an arbitrary normal closed subgroup $N \subset G$ are not-preserved.

\vspace{\baselineskip}
\emph{Acknowledgements.} We would like to thank D. Buchholz and L. Zsido for numerous helpful discussions and suggestions. One of us (G.M.) acknowledges the kind hospitality of the Institute of Theoretical Physics of G\"ottingen University during some stages of this work.

\renewcommand{\thesection}{\Alph{section}}
\setcounter{section}{0}
\section{Some results on nuclear maps}\label{app:nuclear}
For the interested reader, in this appendix we collect some elementary results about nuclear maps between Banach spaces which are used in the main text, but whose proofs are not easily found in the existing literature. We begin by recalling the definition of a nuclear map between Banach spaces.

\begin{definition}\label{def:nuclear}
Let $X, Y$ be Banach spaces and $p \in (0,1]$. A bounded linear map $T \in \sB(X,Y)$ is said to be \emph{$p$-nuclear} if there exist sequences $(f_n)_{n \in \bN} \subset X^*$ and $(y_n)_{n \in \bN} \subset Y$ such that
\begin{gather*}
Tx = \sum_{n=1}^{+\infty} f_n(x)y_n, \qquad \forall x \in X,\\
\sum_{n=1}^{+\infty} \norm{f_n}_{X^*}^p \norm{y_n}_Y^p < +\infty.
\end{gather*}
The \emph{nuclear $p$-norm} of $T$ is defined as
\begin{equation*}
\norm{T}_p := \inf\bigg\{ \bigg(\sum_{n=0}^{+\infty} \norm{f_n}_{X^*}^p \norm{y_n}_Y^p\bigg)^{1/p}\bigg\},
\end{equation*}
where the infimum is taken over all possible decompositions of $T$ as above.
\end{definition}

The $p$-nuclear maps form a vector space equipped with the quasi-norm $\norm{\cdot}_p$~\cite[sec.\ 19.7]{Jarchow:1981a}. A closely related concept is the one of $\eps$-content of a compact map.

\begin{definition}\label{def:content}
Let $X, Y$ be Banach spaces, and $T\in\sB(X,Y)$. The \emph{$\eps$-content} of $T$, denoted by $N_T(\eps)$, is the maximal number of elements $x_i \in X$, $\norm{x_i}\leq 1$, $i =1, \dots, N_T(\eps)$, such that $\norm{T(x_i-x_j)}>\eps$ for $i\neq j$.
\end{definition}

It is easy to check that $N_T(\eps) < +\infty$ for all $\eps > 0$ if and only if $T$ is a compact map. For the convenience of the reader, we summarize in the next lemma some results, which are used in the main body of the paper, about the relationships between $\eps$-content and nuclearity for maps with values in a Hilbert space. For their proof, we refer the reader to~\cite[lemma 2.1]{Buchholz:1996mx} and to the references cited there.

\begin{lemma}\label{lem:contentnuclear}
Let $X$ be a Banach space, $H$ a Hilbert space and $T \in \sB(X,H)$. 
\begin{proplist}{2}
\item \label{part:nucleartocontent} If $0<p<1$ and $q>p/(1-p)$ there exists a constant $c=c_{p,q}>0$ such that, for each $p$-nuclear $T$, there holds\begin{equation}
N_T(\eps) \leq e^{\frac{c\norm{T}_p^q}{\eps^q}}.
\end{equation}
\item \label{part:contenttonuclear} If $0<p\leq 1$ there exists a constant $d=d_p>0$ such that if there exists a sequence of positive numbers $(\eps_m)_{m \in \bN}$ with
\begin{equation*}
\sum_{m=1}^{+\infty}\big(m^\frac{1}{2}\eps_m N_T(\eps_m)^\frac{1}{m}\big)^p < +\infty,
\end{equation*}
then the map $T$ is $p$-nuclear and
\begin{equation}
\norm{T}_p \leq d \bigg(\sum_{m=1}^{+\infty}(m^\frac{1}{2}\eps_m N_T(\eps_m)^\frac{1}{m})^p\bigg)^\frac{1}{p}.
\end{equation}
\end{proplist}
\end{lemma}
 
We will have to deal with tensor products of nuclear maps, and the following lemma will be useful at some instances.

\begin{lemma}\label{lem:tensornuclear}
Let $T_i : X_i \to Y_i$ be $p$-nuclear maps, $i=1,2$, and let $\norm{\cdot}_\a$, $\norm{\cdot}_\b$ be cross-norms on the algebraic tensor products $X_1\otimes X_2$, $Y_1\otimes Y_2$ respectively. Assume further that $\norm{\cdot}_\a$ majorizes the injective cross-norm on $X_1 \otimes X_2$. Then there exists a unique bounded operator $T_1\otimes T_2 : X_1 \otimes_\a X_2 \to Y_1 \otimes_\b Y_2$ such that $T_1\otimes T_2(x_1 \otimes x_2) = T_1(x_1)\otimes T_2(x_2)$, and there holds $\norm{T_1\otimes T_2}_p \leq \norm{T_1}_p \norm{T_2}_p$.
\end{lemma}

\begin{proof}
Uniqueness of $T_1\otimes T_2$ is immediate. We prove existence. For a given $\eps > 0$, we can find sequences $(f_{i,n})_{n \in \bN} \subset X_i^*$, $(y_{i,n})_{n \in \bN} \subset Y_i$, $i=1,2$, such that
\begin{equation*}
T_i(x_i) = \sum_{n=1}^{+\infty} f_{i,n}(x_i) y_{i,n}, \quad \sum_{n=1}^{+\infty} \norm{f_{i,n}}^p\norm{y_{i,n}}^p < (\norm{T_i}_p + \eps)^p.
\end{equation*}
Since $\norm{\cdot}_\a$ majorizes the injective cross-norm, the induced norm $\norm{\cdot}_{\a^*}$ on $X_1^*\otimes X_2^*$ is a cross-norm~\cite[prop. IV.2.2]{Takesaki:1979a} and therefore the algebraic tensor product $f_{1,n}\otimes f_{2,m}$ extends to an element of $X_1^*\otimes_{\a^*}X_2^* = (X_1\otimes_\a X_2)^*$, denoted by the same symbol. Furthermore, since $p \leq 1$, it is easy to check that 
\begin{equation*}
T_1 \otimes T_2(x) = \sum_{n,m = 1}^{+\infty} f_{1,n}\otimes f_{2,m} (x) y_{1,n}\otimes y_{2,m},
\end{equation*}
defines a bounded $T_1 \otimes T_2 : X_1 \otimes_\a X_2 \to Y_1 \otimes_\b Y_2$ such that $T_1 \otimes T_2(x_1\otimes x_2) = T_1(x_1)\otimes T_2(x_2)$, and from
\begin{equation*}
\sum_{n,m = 1}^{+\infty} \norm{f_{1,n}\otimes f_{2,m}}_{\a^*}^p \norm{y_{1,n}\otimes y_{2,m}}_\b^p < (\norm{T_1} + \eps)^p (\norm{T_2} + \eps)^p,
\end{equation*}
and the arbitrariness of $\eps$, we get the estimate $\norm{T_1 \otimes T_2}_p \leq \norm{T_1}_p \norm{T_2}_p$.
\end{proof}

We will apply this result to the case in which the $X_i$ are C$^*$-algebras and $\norm{\cdot}_\a$ is the minimal C$^*$-cross-norm, for which the above hypotheses are satisfied~\cite[sec. IV.4]{Takesaki:1979a}.

\begin{lemma}\label{lem:nuclearbon}
Let $X$ be a Banach space, $H$ be a Hilbert space and $T : X \to H$ a $p$-nuclear map, $0 < p < 2/3$. There exist an orthonormal system $(\xi_n)_{n \in \bN}$ in $\overline{\ran T}$ and a sequence $(f_n)_{n \in \bN} \subset X^*$ such that for each $q$, $4p/(p+2) < q \leq 1$, there holds
\begin{equation*}
Tx = \sum_{n=1}^{+\infty} f_n(x)\xi_n, \qquad \sum_{n=1}^{+\infty} \norm{f_n}^q < +\infty.
\end{equation*}
\end{lemma}

\begin{proof}
The proposition is trivial if $T$ is of finite rank, so we assume that this is not the case. Let $(\z_k)_{k \in \bN} \subset H$, $(g_k)_{k \in \bN} \subset X^*$ be such that
\begin{equation*}
Tx = \sum_{k=1}^{+\infty} g_k(x)\z_k, \qquad \sum_{k=1}^{+\infty} \norm{g_k}^p \norm{\z_k}^p < +\infty,
\end{equation*}
and we can assume that the sequence $a_k := \norm{g_k}\norm{\z_k}$ is non-increasing. If $E$ is the projection on $\overline{\ran T}$, we have $Tx = ETx = \sum_{k=1}^{+\infty} g_k(x)E\z_k$, so that we can also assume that $\z_k \in \overline{\ran T}$, and of course that $\z_k \neq 0$. 

We now define inductively a subsequence $(\e_m)_{m \in \bN} \subset (\z_k)_{k \in \bN}$ of linearly independent vectors in the following way: $\e_1 := \z_1$, and having defined linearly independent vectors $\{\e_1, \dots, \e_m\} \subset (\z_k)_{k \in \bN}$, it will be $\e_m = \z_k$ for some $k$, and $\e_{m+1}$ will be defined to be the first vector in $(\z_h)_{h \geq k+1}$ which is linearly independent from $ \{\e_1, \dots, \e_m\}$. It is clear from this construction that each vector $\z_k$ will be a linear combination of the vectors $\e_1,\dots, \e_k$ at most. 

Let now $(\x_n)_{n \in \bN} \subset \overline{\ran T}$ be the orthonormal system obtained by applying the Gram-Schmidt procedure to $(\e_m)_{m \in \bN}$. It also holds that $\e_m$ is in the subspace spanned by $\x_1, \dots, \x_m$, so that we will have
\begin{equation*}
\z_k = \sum_{n=1}^k \a_{kn}\x_n, \qquad \norm{\z_k}^2 = \sum_{n=1}^k \abs{\a_{kn}}^2,
\end{equation*}
for suitable scalars $(\a_{kn})_{1\leq n \leq k <+\infty}$, and, if $q \leq 1$, it follows, thanks to the convexity properties of the function $t \to t^\frac{q}{2}$, that
\begin{equation}\label{eq:akninequal}
\norm{\z_k}^q = k^{\frac{q}{2}}\bigg(\sum_{n=1}^k \frac{1}{k}\abs{\a_{kn}}^2\bigg)^\frac{q}{2} \geq k^{\frac{q}{2}-1}\sum_{n=1}^k \abs{\a_{kn}}^q.
\end{equation}
Since the sequence $a_k = \norm{g_k}\norm{\z_k}$ is non-increasing, from the inequalities
\begin{equation*}
k a_k^p \leq \sum_{h=1}^k a_h^p \leq \sum_{h=1}^{+\infty}a_h^p < +\infty,
\end{equation*}
we get that there exists a constant $c > 0$ such that $a_k < c/k^{\frac{1}{p}}$. It follows then from this observation and from equation~\eqref{eq:akninequal} that, if we set $\a_{kn} = 0$ for $n > k$,
\begin{equation*}
\sum_{k=1}^{+\infty} \sum_{n=1}^{+\infty}\abs{\a_{kn}}^q\norm{g_k}^q \leq \sum_{k=1}^{+\infty}k^{1-\frac{q}{2}} \norm{\z_k}^q\norm{g_k}^q \leq c^q \sum_{k=n}^{+\infty}k^{1-\frac{q}{2}-\frac{q}{p}}
\end{equation*}
is convergent if $q > 4p/(p+2)$, which entails that $f_n := \sum_{k=n}^{+\infty} \a_{kn}g_k$ is a bounded functional on $X$ with
\begin{equation*}
\sum_{n=1}^{+\infty} \norm{f_n}^q \leq \sum_{n=1}^{+\infty}\sum_{k=1}^{+\infty}\abs{\a_{kn}}^q\norm{g_k}^q 
                            = \sum_{k=1}^{+\infty}\sum_{n=1}^{+\infty}\abs{\a_{kn}}^q \norm{g_k}^q < +\infty,
\end{equation*}
where we have used the fact that if $0 < q \leq 1$ and $a,b >0$ then $(a+b)^q \leq a^q + b^q$, and that it is possible to interchange the sums since the double sum is absolutely convergent. This also implies that for each $x \in X$ the double series $\sum_{k=1}^{+\infty}\sum_{n=1}^{+\infty} \a_{kn}g_k(x) \x_n$ is absolutely convergent in $H$, and therefore it is allowed to interchange sums in
\begin{equation*}
Tx = \sum_{k=1}^{+\infty}g_k(x)\z_k = \sum_{k=1}^{+\infty}g_k(x)\sum_{n=1}^k \a_{kn}\x_n 
   = \sum_{n=1}^{+\infty}\x_n\sum_{k=n}^{+\infty}\a_{kn}g_k(x) = \sum_{n=1}^{+\infty} f_n(x)\x_n,
\end{equation*}
which concludes the proof.
\end{proof}

\providecommand{\bysame}{\leavevmode\hbox to3em{\hrulefill}\thinspace}

\end{document}